\documentclass{elsart}
\usepackage{times}
\usepackage{bm,amsfonts}
\makeatletter
\@ifundefined{pdfoutput} 
{ 
\usepackage[dvips]{graphicx,color}
}
{ 
\ifnum\pdfoutput=0\relax
\usepackage[dvips]{graphicx,color}
\fi
\ifnum\pdfoutput=1\relax
\usepackage[pdftex]{graphicx,color}
\fi
}
\makeatother
\newcommand{\beq}{\begin{equation}}
\newcommand{\eeq}{\end{equation}}
\newcommand{\beqa}{\begin{eqnarray}}
\newcommand{\eeqa}{\end{eqnarray}}
\newcommand{\nn}{\nonumber \\}
\newcommand {\np}[1]{{\mbox{\textrm{:}\,}{#1}{\,\textrm{:}}} }
\def \ra {\rangle}
\def \dd {\mathrm{d}}
\def \e {\mathrm{e}}

\def \el {\textrm{\scriptsize{el}}}

\def \l {\lambda}
\def \la {\langle}
\def \ra {\rangle}
\def \s {\sigma}
\def \t {\tau}
\def \B {{\mathcal B}}

\def \I {{\mathbb I}}

\def \Z {{\mathbb Z}}
\def \ch {\mathrm{ch}}

\def \z {\zeta}
\def \L {\underline{\Lambda}}
\def \la {\langle}
\def \ra {\rangle}
\def \D {\Delta}

\def \PF {\mathrm{PF}}
\def \Im {\mathrm{Im} \, }
\def \mod {\ \mathrm{mod} \ }

\def \uu {{\widehat{u(1)}}}
\def \su {{\widehat{su(3)}}}

\begin{document}
\begin{frontmatter}
\title{Diagonal Coset Approach to Topological Quantum Computation 
with Fibonacci Anyons}
\author{Lachezar S. Georgiev,}
\ead{lgeorg@inrne.bas.bg}
 \author{Ludmil Hadjiivanov \and Grigori Matein}
\address{Institute for Nuclear Research and
Nuclear Energy \\
Bulgarian Academy of Sciences \\
 Tsarigradsko Chaussee 72,  1784 Sofia, BULGARIA}
\begin{abstract}
We investigate a promising conformal field theory realization  scheme for topological quantum computation based on  the
Fibonacci anyons, which are believed to be realized as quasiparticle excitations  in the $\Z_3$ parafermion fractional 
quantum Hall state in the second Landau level with filling factor $\nu=12/5$. These anyons are non-Abelian and are 
known to be capable of universal topological quantum computation. The quantum information is encoded in the fusion channels of 
pairs of such non-Abelian anyons and is protected from noise and decoherence by the topological properties of these systems. 
The quantum gates are realized by braiding
of these anyons. We propose here an implementation of the $n$-qubit topological quantum register in terms of $2n+2$ 
Fibonacci anyons. The matrices emerging from the anyon exchanges, i.e. the \textit{generators} of the braid group for one qubit are 
derived from the coordinate wave functions of a large number of electron holes and 
4 Fibonacci anyons which can furthermore be represented as correlation functions in $\Z_3$ parafermionic two-dimensional conformal field theory.
The representations of the braid groups for more than 4 anyons are obtained by fusing pairs of anyons before braiding,
thus reducing eventually the system to 4 anyons.
\end{abstract}
\end{frontmatter}
\section{Introduction}
Quantum technologies have recently experienced a second quantum revolution \cite{QTech,QTech-book}. Quantum computers \cite{nielsen-chuang} 
in particular are expected 
soon to become more powerful compared to classical supercomputers and quantum supremacy has been clearly demonstrated in the 
last 5 years \cite{supremacy-2019,supremacy-2021}. Yet, stable and resilient realizations of qubits and exact execution of the quantum gates are still challenging and 
insufficient for solving practical physical or mathematical problems. 

Topological Quantum Computation (TQC) is an attractive paradigm \cite{kitaev-TQC,preskill-TQC,sarma-RMP,simon-TQ} 
for quantum computing which uses degenerate multiplets of 
quantum states to implement  the multi-qubit registers.  Quantum information is encoded in the fusion channels of non-Abelian 
anyons \cite{wilczek}  and fusion paths Bratteli diagrams \cite{sarma-RMP}.
Quantum gates are then executed by braiding (i.e., exchanging clockwise and anti-clockwise) 
 these non-Abelian anyons. Both the information encoding and the quantum gates execution are 
topologically protected from noise which makes TQC very promising.

The most studied TQC scheme is that of the Ising anyons realized as quasiholes in the Moore--Read Pfaffian Fractional Quantum Hall (FQH) state \cite{mr} 
believed to be observed \cite{5-2-book} at filling factor $\nu=5/2$. This FQH state is the most stable one among all states that are expected to possess non-Abelian 
quasiparticle excitations  \cite{5-2-book}. However, the Pfaffian FQH state is not capable of universal TQC, since the representation of the Braid group contains only Clifford 
gates which form a finite subgroup of $SU(2^n)$ for $n$ qubits \cite{sarma-RMP,clifford}.

Another FQH state observed experimentally at filling factor $\nu=12/5$ in the second Landau level \cite{choi-west-08}, has also attracted much attention. 
Analytical and numerical arguments suggest that this state might be  described by the $k=3$ Read--Rezayi state \cite{rr} and is believed to possess 
non-Abelian quasiparticles, which are topologically equivalent to the Fibonacci anyons \cite{preskill-TQC}. 
The braiding matrices of the Fibonacci anyons, which can be realized in a diagonal CFT coset \cite{NPB2001}, are non-diagonal and are expected to be 
universal for topologically protected quantum information processing \cite{freedman-larsen-wang-TQC}. 
However, the energy gap of this FQH state is an order of magnitude lower \cite{choi-west-08} than that of the FQH state observed at filling factor $\nu=5/2$.
This increases drastically the noise/signal ratio at $\nu=12/5$, respectively the stability decreases. Nevertheless, the braiding properties of the Fibonacci anyons, 
realized by triples of fundamental quasiholes of the $k=3$ Read--Rezayi state, have been extensively investigated obtaining remarkable 
results \cite{bonesteel-2005,Hormozi-TQC,Hormozi-TQC-RR}.

In this paper we propose another implementation of the Fibonacci TQC by using the $\varepsilon$ primary field of  the $k=3$ parafermion FQH state realized 
as a diagonal CFT coset \cite{NPB2001}.
This scheme is  similar to the TQC scheme of Das Sarma et al. \cite{sarma-freedman-nayak},
 which was originally based on monodromy transformations of the Pfaffian wave functions for the Ising Conformal Field Theory (CFT) model.
This is done in such a way
to construct by braiding the single-qubit Hadamard and phase gates as well as the two-qubit  Controlled-Z and
Controlled-NOT gates \cite{TQC-PRL,TQC-NPB}. These gate constructions are naturally topologically protected from noise and decoherence. 

Compared to the existing TQC models the proposed new topological quantum computation scheme with Fibonacci anyons realized by 
the $\varepsilon$ field in the $\Z_3$ parafermion FQH state is:
\begin{itemize}
\item 
\textbf{Natural:} 
TQC assumes that the different basis vectors in the single-qubit space are realized by the different fusion channels of pairs of anyons. Since there are 
only 2 fusion channels for the Fibonacci anyons, one qubit is essentially one pair  of anyons, just like in the Moore--Read FQH state where the Ising anyons 
also have 2 fusion channels. \\
\item
\textbf{Simpler:} The qubit initialization scheme proposed in this paper is similar to but actually simpler than the Ising anyons TQC scheme of \cite{sarma-freedman-nayak}. 
As we will see in Sect.~\ref{sec:Z_3} the diagonal coset approach 
to the Fibonacci TQC described in Ref.~\cite{NPB2001} is more appropriate than the approach 
using the $\widehat{su(3)}_2/u(1)$ because the former naturally inherits the $\Z_3$ parafermion pairing rule given in Eq.~(\ref{PR}) from its Abelian parent as 
shown in Ref.~ \cite{NPB2001}. \textit{It is this rule that allows us to 
initialize the Fibonacci TQC using the $\varepsilon$ fields localized on antidots threaded by one quantum of magnetic flux}. \\
\item
\textbf{Most economic:} Only $2n+2$ Fibonacci anyons are needed to construct an $n$-qubit register. Just like in the Ising TQC scheme one pair of anyons is inert
(or, as in our case, the first and the last anyons are inert)  and serve to
compensate the total topological charge of the $n$-qubit register. This way the CFT correlation function is non-zero. It is obvious that $2n+2$ is the smallest number 
of  Fibonacci anyons needed to construct $n$ qubits. The Preskill TQC scheme requires $4n$ Fibonacci  anyons for $n$ qubits \cite{preskill-TQC}. 
Other schemes using fundamental quasiholes require $3n$ quasiholes to represent $n$ qubits \cite{bonesteel-2005,Hormozi-TQC}. \\
\item
\textbf{Most efficient:} The braid matrices in this approach have the smallest dimension compared to the other Fibonacci anyons models. For example, the braid matrices for 
two qubits, which are realized by 6 Fibonacci anyons have dimension 5 and those for 3 qubits, realized by 8 anyons have dimension 13. Most of the braid generators
have a direct sum structure, in an appropriate basis, which makes their explicit derivation almost obvious \cite{HG-Monodromy}. This allows us to write the 
generators of the braid 
group $\B^{(n)}$ for Fibonacci anyons for general $n$ which are given in  \cite{HG-Monodromy}.\\
\item
\textbf{Universal:} 
The quantum gates realized by braiding of Fibonacci anyons are known to be capable of universal topological quantum computation in the sense that any 
quantum gate can be approximated with arbitrary precision by consecutive braids.
\end{itemize}

\textit{Summary of results}: 
In Sect.~\ref{sec:Fibonacci} we present general information about the Fibonacci anyons, their Bratteli diagrams and the possibilities to construct qubits.
A short introduction to the CFT realization of the $\Z_3$ parafermion  (Read--Rezayi) FQH state is given in Sect.~\ref{sec:Z_3}. 
Sect.~\ref{sec:single-qubit} discusses how to use the CFT four--quasihole wave functions to construct the elementary qubit's wave-function using the 
occupation number representation and CFT primary operators. Following the wave-function construction \cite{NPB2001} 
as symmetrization of Abelian CFT correlation functions and the explicit computation in terms of hypergeometric functions \cite{ardonne-schoutens} we summarize the 
computation, explicitly given in \cite{HG-Monodromy}, of the 4-Fibonacci anyons wave function with $N=3r$ electron fields where $r$ is a positive integer.
In Sect.~\ref{sec:fibonacci-Z_3} we describe the physical realization of the Fibonacci anyons in the $\Z_3$ parafermion  FQH state by the Laughlin argument to 
localize $\varepsilon$ fields on antidots threaded by one flux quantum as well as their interferometric measurements.
Sect.~\ref{sec:braid} summarizes the generators of the braid group $\B^{(4)}$ for 4 Fibonacci anyons and one qubit derived in \cite{HG-Monodromy} and we construct
explicitly several approximations to the most widely used single-qubit gates.
In Sect.~\ref{sec:Two-qubit} we describe the computational basis for two qubits realized by Bratteli diagrams for 6 Fibonacci anyons and summarize the 
computation, explicitly given in \cite{HG-Monodromy}, of   the generators of the braid group $\B^{(6)}$. Then, we show how they can be used to construct 
various two-qubit gates.
In Sect.~\ref{sec:Three-qubit} we describe the computational basis for three qubits realized by Bratteli diagrams for 8 Fibonacci anyon 
(separating the 8 computational-basis states from the 5 non-computational states) and summarize the 
computation of   the generators of the braid group $\B^{(8)}$. How these can be used to construct various three-qubit gates is also shown.
\section{The Fibonacci anyons and universal TQC: an overview}
\label{sec:Fibonacci}

The $\Z_3$ parafermion FQH state is one of the few physical systems in which the Fibonacci anyons are believed to be experimentally 
 realizable.

The Fibonacci anyon can be identified with the $\Z_3$ parafermion  primary field of CFT dimension $2/5$, denoted by $\varepsilon$
in Tab.~\ref{tab.7}
\[
\varepsilon \equiv \Phi(\L_1+\L_2), \quad \varepsilon^*=\varepsilon, \quad \D_{\varepsilon}=\frac{2}{5},
\]
which is self-dual with respect to the $\Z_3$ parafermion charge conjugation, i.e., it coincides with its antiparticle and can therefore annihilate with 
itself fusing to the vacuum $\I$.


The fusion rules in which will be interested for TQC are those for the two fields $\I$ and $\varepsilon$ given in Tab.~\ref{tab-f}
\beq \label{fusion-fibonacci}
\I \times \I = \I , \quad \I \times \varepsilon = \varepsilon, \quad \varepsilon \times \varepsilon = \I +\varepsilon .
\eeq
The quantum dimension of the Fibonacci anyon $\varepsilon$ is the Golden ratio $d_{\varepsilon} = \delta=\frac{1+\sqrt{5}}{2}$ 
which follows from the last fusion rule since $d_{\varepsilon}^2 = d_{\varepsilon} +1$, or, alternatively, from the definition of the quantum 
dimension in terms of modular $S$ matrix \cite{CFT-book,JGSP-S-PF}, i.e., $d_{\varepsilon}=S_{0\varepsilon}/ S_{00}$, with $0$ 
labeling the vacuum sector fo the $\Z_3$ rational CFT.

\begin{figure}[htb]
	\centering
	\includegraphics[viewport=30 330 560 510,clip,width=13cm]{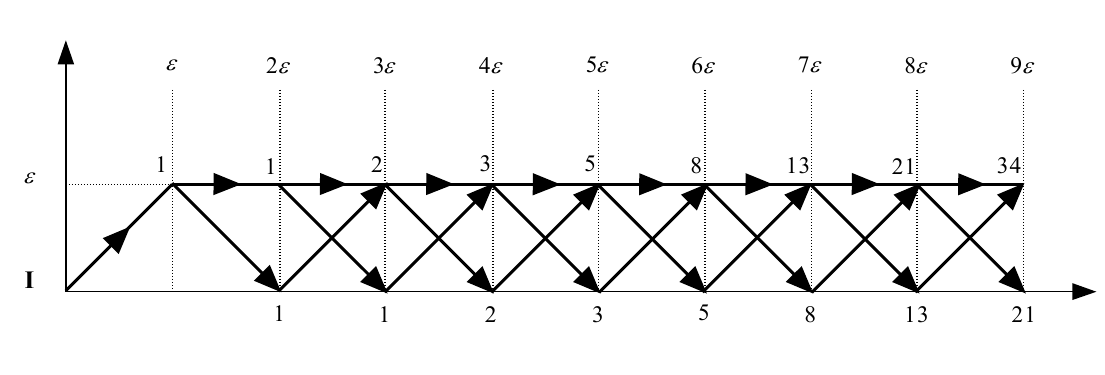} 
\caption{Bratteli diagram for Fibonacci anyons. Each step to the right adds by fusion one Fibonacci anyon $\varepsilon$. 
On the horizontal axis we mark the vacuum $\I$ (denoted by \textbf{I} in the figure), as the result of the fusion step according to Eq.~(\ref{fusion-fibonacci}), 
while on the horizontal dashed line labeled by $\varepsilon$ we mark the Fibonacci anyon as the result of the fusion step. The numbers next
to the vertical dashed lines are the Fibonacci numbers (see the text below).
 \label{fig:brattelli-fibonacci}}
\end{figure}

The multiparticle quantum states of Fibonacci anyons are labeled by fusion paths which are conveniently depicted on the Bratteli diagram
shown in Fig.~\ref{fig:brattelli-fibonacci}. 
If we start from the vacuum $\I$, which is depicted in Fig.~\ref{fig:brattelli-fibonacci} by $1$ and add
one Fibonacci anyon the result is a single Fibonacci anyon $\varepsilon$. This fusion step is denoted by the first arrow in the diagram, 
which is at $45^{\circ}$. Fusing a second $\varepsilon$ produces two channels: the vacuum channel, for which the arrow points to 
$-45^{\circ}$ and the $\varepsilon$ channel, for which the arrow is horizontal pointing to the right. Adding a third  $\varepsilon$ 
starting from lower channel representing the vacuum on the horizontal axis produces one arrow at $45^{\circ}$, while starting from the 
$\varepsilon$ channel of the fusion of the two previous $\varepsilon$ produces two more arrows to the right: one pointing at $-45^{\circ}$
and one horizontal arrow. The numbers written close to the vertical dashed lines representing the number $n\varepsilon$ of Fibonacci anyons 
give the number of orthogonal fusion-path states with definite topological charge, i.e., the numbers close to the  horizontal axis correspond 
to the number of orthogonal states with $n$ of Fibonacci anyons and total topological charge $1$ (the vacuum), while the numbers 
 close to the  horizontal dashed line denoted by $\varepsilon$ correspond  to the number of orthogonal states with $n$ of Fibonacci anyons 
and total topological charge $\varepsilon$. It is interesting to note that the multi-anyon state of $n+1$ of Fibonacci anyons with trivial 
topological charge $\I$ can be obtained in only one way--by adding one $\varepsilon$ to the multi-anyon state of $n$ of Fibonacci anyons 
with total topological charge $\varepsilon$. That is why the numbers of states on the two ends of the $-45^{\circ}$ arrows are always the same.
On the other hand, the multi-anyon state of $n+1$ of Fibonacci anyons with total topological charge $\varepsilon$ can be obtained in two 
ways--one, by adding one $\varepsilon$ to the multi-anyon state of $n$ of Fibonacci anyons 
with total topological charge $\varepsilon$ and second, by   adding one $\varepsilon$ to the multi-anyon state of $n$ of Fibonacci anyons 
with total topological charge $\I$. Therefore, the dimension $D_{n+1}^{(\varepsilon)}$  of the orthogonal fusion-path states with 
$n+1$ of Fibonacci anyons with total topological charge $\varepsilon$  is
\beq \label{D_n-e}
D_{n+1}^{(\varepsilon)} = D_n^{(\varepsilon)} + D_{n-1}^{(\varepsilon)}, \quad D_1^{(\varepsilon)} = D_{2}^{(\varepsilon)}=1,
\eeq
while the dimension $D_{n+1}^{(\I)}$  of the orthogonal fusion-path states with $n+1$ of Fibonacci anyons with total topological 
charge $\I$  is $D_{n+1}^{(\I)} = D_n^{(\varepsilon)}$, i.e., 
\beq \label{D_n-1}
D_{n+1}^{(\I)} = D_n^{(\I)} + D_{n-1}^{(\I)}, \quad D_2^{(\I)} = D_{3}^{(\I)}=1.
\eeq
 Because Eqs.~(\ref{D_n-e}) and (\ref{D_n-1}) are the recursion relations for the Fibonacci numbers, these dimensions 
explain the origin of the name of the Fibonacci anyons.

There is an explicit formula for the Fibonacci numbers (\ref{D_n-e}), which can be easily proved by mathematical induction,
\[
D_n^{(\varepsilon)} =: D_n= \frac{1}{\sqrt{5}} \left[  \left(   \frac{1+\sqrt{5}}{2}\right)^n  -  \left( \frac{1-\sqrt{5}}{2}\right)^n   \right] 
\]
which gives a more physical meaning of the word ``quantum'' dimension. For large $n$ the Fibonacci numbers can be approximated by
\[
 D_n \mathop{\simeq}\limits_{n \gg 1} \frac{1}{\sqrt{5}} \left(   \frac{1+\sqrt{5}}{2}\right)^n = \mathrm{const} \times (d_\varepsilon)^n
\]
because the second term is subleading. Then the physical interpretation is that the dimensions of the Hilbert space  spanned by 
the multianyon wave function corresponding to $n$ Fibonacci anyons at fixed positions is simply proportional to the quantum
 dimension $d_\varepsilon = \delta$ to the power $n$ \cite{preskill-TQC}.

\begin{figure}[htb]
	\centering
	\includegraphics[viewport=30 300 560 480,clip,width=13cm]{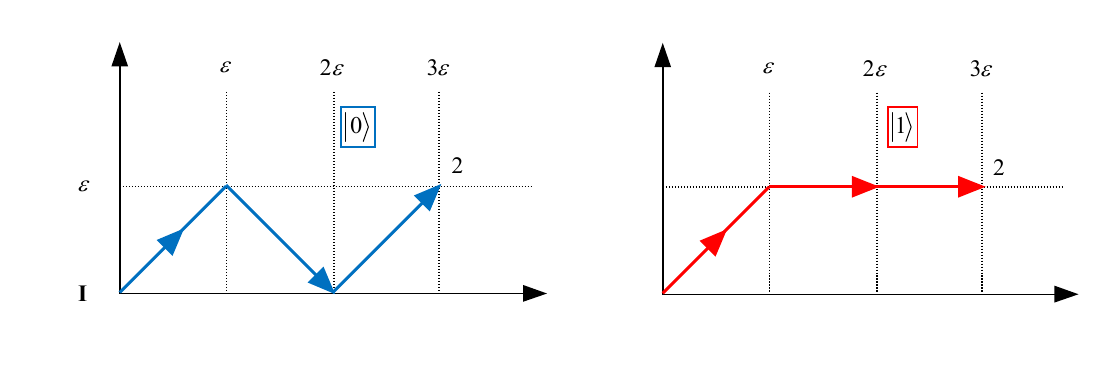} 
\caption{Bratteli diagrams for the two states in the computational basis  $\left\{  |0\ra ,  |1\ra \right\}$ of $3$ Fibonacci anyons with total 
topological charge $\varepsilon$. 
 \label{fig:3-fibonacci}}
\end{figure}

\begin{figure}[htb]
	\centering
	\includegraphics[viewport=30 340 560 490,clip,width=13cm]{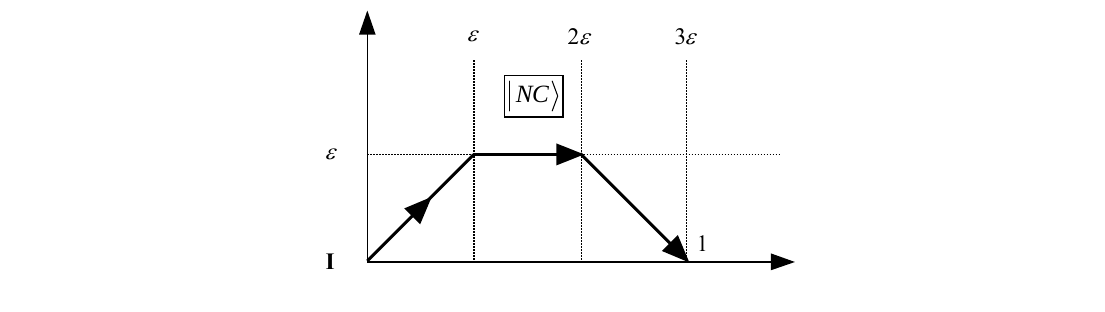} 
\caption{Bratteli diagram for the non-computational state $|NC\ra$ with 3 Fibonacci anyons and trivial total topological charge $\I$ 
(denoted by \textbf{I} in the figure). 
 \label{fig:3-fibonacci-NC}}
\end{figure}

\begin{figure}[htb]
	\centering
	\includegraphics[viewport=30 330 560 510,clip,width=13cm]{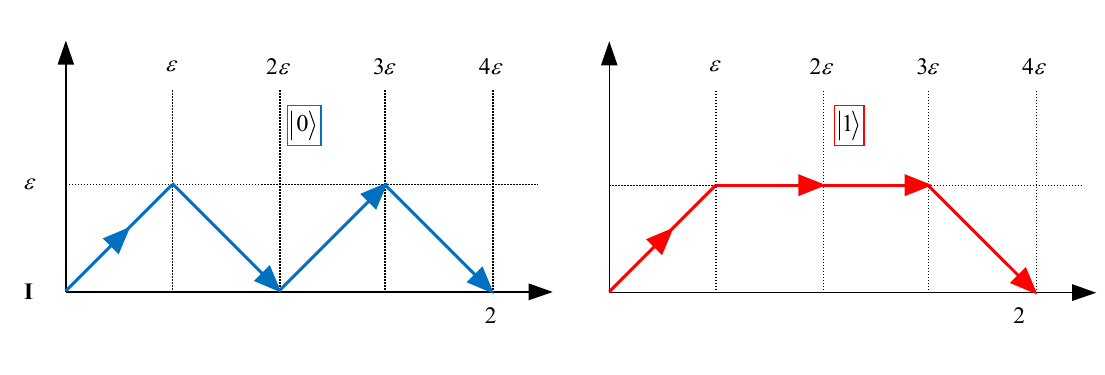} 
\caption{Bratteli diagrams for the computational basis  $\left\{  |0\ra ,  |1\ra \right\}$ of $4$ Fibonacci anyons with total 
topological charge $\I$ (denoted by \textbf{I} in the figure).  \label{fig:4-fibonacci}}
\end{figure}

In order to construct a qubit with Fibonacci anyons we need to identify a two dimensional space of orthogonal fusion-path states and 
the quantum information encoding  is again in the fusion channels and fusion paths. Looking in Fig.~\ref{fig:brattelli-fibonacci} 
we see that there are two different ways to construct a two-dimensional space: $ D_3^{(\varepsilon)}=2$ or  $D_{4}^{(\I)} =2$,
i.e., $3$ Fibonacci anyons with  total topological charge $\varepsilon$, as shown in  Fig.~\ref{fig:3-fibonacci}, or  $4$ Fibonacci anyons 
with  trivial total topological charge $\I$ as shown in  Fig.~\ref{fig:4-fibonacci}.
We have to emphasize here that there is one more fusion-path state,  that can be realized with $3$ Fibonacci anyons, which has a 
trivial total  topological charge $\I$ (denoted by \textbf{I} in the figure)
\cite{preskill-TQC,bonesteel-2005,sarma-RMP}, as shown in Fig.~\ref{fig:3-fibonacci-NC} and is called a non-computational state $|NC\ra$.
This state actually decouples from the two states with total  topological charge $\varepsilon$ because $|NC\ra$ belongs to 
 a different superselection sector.
For this reason all braiding matrices decouple as direct sums for the states $\left\{  |0\ra ,  |1\ra \right\}$ and $|NC\ra$ as we shall see later.
We can write symbolically the fusion channel between two Fibonacci anyons as a subscript of the pair $(\varepsilon,\varepsilon)_{\I}$ 
and $(\varepsilon,\varepsilon)_{\varepsilon}$. Then the  fusion path for $3$ Fibonacci anyons can be written as 
$((\varepsilon,\varepsilon)_{\varepsilon}, \varepsilon)_{\I}$ and $((\varepsilon,\varepsilon)_{\varepsilon}, \varepsilon)_{\varepsilon}$.
Now the two computational states and the non-computational state ca be written as  \cite{bonesteel-2005}
\beqa
|0\ra &=&((\varepsilon,\varepsilon)_{\I} , \varepsilon)_{\varepsilon}  \nn
|1\ra &=&((\varepsilon,\varepsilon)_{\varepsilon}, \varepsilon)_{\varepsilon}  \nn
|NC\ra &=&((\varepsilon,\varepsilon)_{\varepsilon}, \varepsilon)_{\I} \nonumber .
\eeqa
Another way to represent a qubit, which is more appropriate for the CFT implementation, is to use 4 Fibonacci anyons with total topological charge $\I$. 
The dimension of the Hilbert space is 2 as can be seen from Fig.~\ref{fig:brattelli-fibonacci}.
 In that case the computational basis $\left\{  |0\ra ,  |1\ra \right\}$
of the fusion-path states is shown in Fig.~\ref{fig:4-fibonacci} and can be written symbolically as CFT correlation functions with fixed 
fusion channels and fusion paths corresponding to the Bratteli diagrams shown in Fig.~\ref{fig:4-fibonacci}
\beqa \label{4F-comp-basis}
|0\ra &=&\la (((\varepsilon,\varepsilon)_{\I} , \varepsilon)_{\varepsilon}, \varepsilon)_{\I}  \ra  = 
\la (\varepsilon,\varepsilon)_{\I} , (\varepsilon, \varepsilon)_{\I}  \ra\nn
|1\ra &=&\la (((\varepsilon,\varepsilon)_{\varepsilon}, \varepsilon)_{\varepsilon}, \varepsilon)_{\I}\ra  =
\la (\varepsilon,\varepsilon)_{\varepsilon} , (\varepsilon, \varepsilon)_{\varepsilon}  \ra  ,
\eeqa
where we used the associativity of the fusion rules which allows to fuse the first and the last pair of anyons and then to fuse further 
the resulting anyons. For example, in the state $|1\ra$ the first two Fibonacci anyons fuse through the channel
$\varepsilon \times \varepsilon \to \varepsilon$ and so do the last two Fibonacci anyons. 
After the fusion of the first and the last pairs of anyons the resulting two $\varepsilon$ inside the CFT correlation function in the state 
$|1\ra$  in Eq.~(\ref{4F-comp-basis}) fuse further to $\I$ because 
the second channel gives zero contribution due to $\la\varepsilon \ra =0$ which holds in the CFT. 
One advantage of this representation of a qubit is that there is no non-computational state and therefore no leakage out of the single-qubit's Hilbert space. 
We will consider in  more detail the CFT realization of the single qubit in Sect.~\ref{sec:single-qubit} below.
This pair representation of the qubits is similar to that in the Ising model where quantum information is encoded in the fusion channel 
of the first pair of Ising anyons, while the last pair of anyons is inert  (or, vice versa). This means that it carries no information and only ensures that the CFT 
correlation function is not zero \cite{TQC-NPB}.
%
\section{The $\Z_3$ parafermion fractional quantum Hall state}
\label{sec:Z_3}
The $\Z_3$ parafermion  (Read--Rezayi) state is the most promising candidate to describe the experimentally 
observed \cite{pan,xia,pan-xia-08} incompressible state in the second Landau level, corresponding to filling factor 
  $\nu_H= 12/5 = 3-3/5$. The most appealing characteristics of this FQH state is that it possesses non-Abelian
quasiparticle excitations, which are topologically equivalent  to the Fibonacci anyons \cite{bonesteel-2005,sarma-RMP}. These anyons  can be used to 
construct multielectron wave functions which belong to degenerate manifolds (of wave functions with electrons and anyons at fixed positions) 
whose dimension increases exponentially with the number 
of anyons and can be used for \textit{universal} topologically protected quantum information processing \cite{bonesteel-2005,sarma-RMP}.
While the $\nu_H=12/5$ FQH state is less stable than the more popular $\nu_H=5/2$ FQH state, which is believed to be described by
the Majorana fermion of the Ising model, the latter is known to be not universal \cite{clifford}, in the sense that not all elementary 
quantum operations can be 
implemented by braiding Ising anyons \cite{TQC-NPB,sarma-RMP} and are therefore not topologically protected. 
On the contrary, all elementary quantum gates in the Fibonacci quantum computer can be implemented
by braiding of Fibonacci anyons \cite{bonesteel-2005} and all they are fully protected from noise and decoherence by 
the topology of the quantum computer \cite{sarma-RMP}. 

The rational CFT describing the $\Z_3$ parafermion FQH state has been constructed \cite{NPB2001} as a diagonal coset 
of a parent Abelian CFT, whose charges form a maximally symmetric chiral quantum Hall lattice in the terminology of \cite{fro-stu-thi} and can be
represented as a direct sum  
\beq\label{PF_3}
\left(\uu_{15} \oplus \PF_3\right)^{\Z_3}, \quad 
\PF_3= 
\frac{\widehat{su(3)}_1\oplus \widehat{su(3)_1}}{\widehat{su(3)_2}}.
\eeq
 Here $\uu_{15}$ is the $u(1)$ current algebra, rationally extended with a pair of vertex exponents 
$\np{\exp\left(\pm i \sqrt{15} \phi(z)\right)}$
 of a normalized chiral boson $\phi(z)$, i.e.,
\beq\label{normal}
\la \phi(z)\phi(w) \ra = -\ln(z-w) .
\eeq 
This $u(1)$ algebra represents the electric properties of the 
edge excitations in the FQH liquid.
 The neutral part $\PF_3$ is a diagonal coset construction with Virasoro central charge $4/5$  \cite{NPB2001}  of the current algebra $\su_{1}\oplus \su_{1}$ 
factorized by its diagonal subalgebra
 $\su_{2}$ generated by the sums of the generators of the numerator in $\PF_3$.  The neutral sector has no contribution to the electric properties but 
describes other topological properties of the edge excitations, such as the statistical angle, fusion and braiding of particle-like excitations.
\begin{table}[htb]
\begin{center}
\begin{tabular}{| c  ||  c | c | c | c | c |}
\hline
$\L_\mu + \L_\nu$ & $\D(\L_\mu + \L_\nu)$ & $P$ & $\sigma$ & $Q$ &  $\Z_3$ Field \\
\hline\hline
$\L_0 + \L_0$   &  $0$ & $0$   & $0$ & $0$ & $ \I$ \\
\hline
$\L_0 + \L_1$ & $\frac{1}{15}$ & $1$  & $1$ & $1$ &  $\sigma_1$ \\
\hline
 $\L_0 + \L_2$ & $\frac{1}{15}$ & $2$ & $2$ & $2$ & $\sigma_2$  \\
\hline
$\L_1 + \L_1$  & $\frac{2}{3}$ & $2$   & $0$ & $1$ &   $\psi_1$\\
\hline
$\L_1 + \L_2$  & $\frac{2}{5}$ & $0$ & $1$ & $2$ & $\varepsilon$ \\
\hline
 $\L_2 + \L_2$  & $\frac{2}{3}$ & $1$ & $0$ & $2$ & $\psi_2$\\
\hline
\end{tabular}
\end{center}
\caption{\label{tab.7}Quantum numbers for the diagonal coset (\ref{PF_3}): the weight of the coset primary  field, its conformal dimension $\Delta$, 
$\Z_3$ charge $P= \mu +\nu \mod 3$, $\sigma$, $Q$ and field notation.}
\end{table}
The $\Z_3$ superscript in Eq.~(\ref{PF_3}) represents the $\Z_3$-pairing rule for combining charged and neutral excitations. 
This is a general pairing rule which applies for all FQH states' CFTs for which the numerator of the filling factor is $n_H > 1$ \cite{LT9}.
In our case the $\Z_3$ pairing rule is simply
\beq \label{PR}
\mu+\nu=l \mod 3
\eeq
where $\L_\mu + \L_\nu$ labels the parafermion primary field given in Table~\ref{tab.7} and $l \mod 5$ labels the $\uu$ primary field 
$\np{\exp\left( i \frac{l}{\sqrt{15}} \phi(z)\right)}$.

The full chiral partition functions of the $\Z_3$ parafermion FQH state is given by \cite{NPB2001}
\beq \label{full-ch}
\chi_{l,\rho} (\t,\z) = \sum_{s=0}^{2} K_{l+5s}(\t,3\z;15) \ch(\L_{l-\rho+s}+\L_{\rho+s})(\t) \quad
\eeq
and describes the spectrum of the edge excitations of the different topological sectors of the CFT (\ref{PF_3}).
Here the charge label is defined $l \mod 5 $ and the neutral label is defined $\rho \mod 3$ with the 
natural restriction $l-\rho \leq \rho$ \cite{NPB2001,NPB2015-2}.
The characters of the irreducible representations of the diagonal coset (\ref{PF_3}) labeled by the weight $\L_{\mu}+\L_{\nu}$
are denoted as   $\ch(\L_{\mu}+\L_{\nu})(\t)$
and $K_l(\t,\z;m)$ are the $\uu$ partition functions   for the charged part which is completely determined by the 
filling factor $\nu_H$ and 
coincides with that for a chiral Luttinger liquid with a compactification radius \cite{cz} $R_c=1/m$, in the 
notation of  \cite{CFT-book,NPB-PF_k} 
 \beq \label{K}
K_{l}(\t,\z; m) = \frac{\mathrm{CZ}}{\eta(\t)} \sum_{n=-\infty}^{\infty} q^{\frac{m}{2}\left(n+\frac{l}{m}\right)^2} 
\e^{2\pi i \z \left(n+\frac{l}{m}\right)}.
\eeq
Here  $q=\e^{2\pi i \t }=\e^{-\beta\Delta\varepsilon}$, where $\beta=(k_B T)^{-1}$ is the inverse temperature,  $ \Delta\varepsilon=h  v_c/L $ is the non-interacting 
energy spacing on the edge,
\[
\eta(\t)=q^{1/24}\prod_{n=1}^\infty (1-q^n)
\]
 is the Dedekind function \cite{CFT-book} and 
 $\mathrm{CZ}(\t,\z)=\exp(-\pi\nu_H(\Im \z)^2/\Im\t)$ is the Cappelli--Zemba factor  needed to preserve the 
invariance of $K_{l}(\t,\z; m)$ with respect to the Laughlin spectral flow \cite{cz}.

The modular parameter $\z$ used in the definition of the rational CFT partition functions is related to the chemical potential $\mu_c$
by \cite{NPB-PF_k,NPB2015}
\beq \label{zeta}
\z= \frac{\mu_c}{\Delta\varepsilon} \t, \quad \mathrm{where} \quad  \tau=i\pi\frac{T_0}{T}, \quad T_0=\frac{h v_c}{\pi k_B L}
\eeq
and transforms after introducing Aharonov--Bohm flux $\phi$ as \cite{NPB2001,NPB2015}
\beq \label{zeta-phi}
\z \to \z+\phi \t \quad \Longleftrightarrow \quad \mu_c \to \mu_c +\phi \Delta \varepsilon.
\eeq
The neutral partition functions  $\ch(\L_{\mu}+\L_{\nu})(\t)$ of the diagonal coset $\PF_3$ are labeled 
in principle by an admissible weight for the current algebra $\widehat{su(3)}_2$, which can be written as a sum of two 
fundamental $su(3)$ weights, i.e., $\L_{\mu}+\L_{\nu}$ with $0\leq \mu \leq \nu \leq 2$.
Following \cite{NPB2001} (and the references therein) these characters for the diagonal coset CFT can be written as
\beq\label{5.18}
\ch_{\s,Q}(\t;\PF_3)= q^{ \D^\PF(\s) - \frac{1}{30}  }
\sum\limits_{
\mathop{m_1,m_2=0}\limits_{ m_1+ 2 m_2 \equiv Q \mod 3} }^\infty
\frac{q^{\underline{m}. C^{-1}.
\left(\underline{m} - \L_\s \right)} }{(q)_{m_1}  (q)_{m_2} },
\eeq
where $(q)_n=\prod\limits_{j=1}^n (1-q^j)$ , $\underline{m} = (m_1, m_{2})$ is a $2$ component vector with 
non-negative integer components  in the basis of $su(3)$ fundamental weights $\{ \L_1, \L_{2}  \}$,
$\D^\PF(\L_\s)=\s(3-\s)/30$ is the CFT dimension of the primary field characterized by the coset triple  \cite{NPB2001} 
of weights
$(\L_\s,0;\L_0+\L_\s)$, for $\L_\s \in \{0,\L_1,\L_{2} \}$ and $C^{-1}$ is the  inverse $su(3)$ Cartan matrix  \cite{CFT-book,NPB-PF_k}.

The parafermion characters $\ch_{\s,Q}\equiv \ch(\L_\mu+\L_\nu)$ derived in Eq.~(\ref{5.18}) are the true 
characters \cite{NPB2001} of the diagonal coset $\PF_3$  labeled  in the standard way by the level-2 weights
$\L_\mu+\L_\nu$, where $0\leq \mu \leq \nu \leq 2$. Then the parameters $(\s,Q)$ are related to $(\mu, \nu)$ by \cite{NPB2001}
\beq \label{mu-rho}
\s=\nu-\mu, \quad Q=\nu  \quad \Longleftrightarrow \quad
\mu=Q-\s, \quad  \nu= Q .
\eeq
The partition functions given in Eqs.~(\ref{full-ch}), (\ref{K}) and (\ref{5.18}) will be used in Sect.~\ref{sec:ini}  to identify the quasiparticles which will be 
localized on the antidot's edge in response to threading the andtidot with one quantum of Aharonov--Bohm flux and this will be the initialization 
procedure for the Fibonacci TQC.
The fusion rules for the $\Z_3$ parafermion fields are given in Tab.~\ref{tab-f}.
\begin{table}[htb]
\begin{center}
\begin{tabular}{lllll}
$\s_1\times  \s_1 =  \s_2 \oplus \psi_1$ & & $\s_2\times \s_2 =\s_1 \oplus \psi_2$ & &
$\psi_1 \times \psi_1=\psi_2$
\\
$\s_1\times \s_2 = \I \oplus  \varepsilon$        & & $\s_2\times \psi_1 =  \s_1$    & &
$\psi_1\times \varepsilon =  \s_2$
\\
$\s_1\times \psi_1 = \varepsilon$         & & $\s_2\times \varepsilon =  \s_2 \oplus  \psi_1$ & &
$\psi_1\times \psi_2 = 1 $
\\
$\s_1\times \varepsilon = \s_1 \oplus \psi_2$  & & $\s_2\times \psi_2 =  \varepsilon $      & &
$\varepsilon \times \varepsilon = \I \oplus \varepsilon$
  \\
$\s_1\times \psi_2 = \s_2$       & &  $\psi_2\times \psi_2=\psi_1$  & &
$\varepsilon \times \psi_2 = \s_1$ 
\end{tabular}
\end{center}
\caption{Fusion rules for the primary fields defined in Table~\ref{tab.7} within the diagonal-coset realization of the $\Z_3$ 
parafermions. \label{tab-f}}
\end{table}

\section{Four--quasihole wave functions and the elementary qubit}
\label{sec:single-qubit}
%
The universality classes of the FQH states,  including their topological properties, can be well described by the effective
 CFT for the edge excitations in the thermodynamic limit \cite{fro-stu-thi,cz,mr}. Since we intend to obtain the elementary
 braid matrices by explicitly exchanging the coordinates of the Fibonacci anyons we need to find the coordinate wave functions
 for a number of anyons and a (big) number $N$ of electrons representing the incompressible FQH liquid. To this end we need to realize the 
quantum states with many Fibonacci anyons as  correlation functions of the $\Z_3$ parafermion CFT.  

We start with the ground state of the  $\Z_3$ parafermion FQH state, i.e, the state with $N$ electrons but without quasiparticle excitations, which can be written 
as \cite{rr,NPB2001}
\[
\Phi_{\mathrm{GS}}(z_1, \ldots, z_N)=\la Q_{\mathrm{bg}} |  \prod_{i=1}^{N} \psi_{\mathrm{hole}}(z_i) | 0 \ra \ 
\exp\left( -\frac{1}{4} \sum_{j=1}^N |z_j|^2\right)
\]
where the exponent represents the standard Gaussian factor for the Landau problem (we set the magnetic length $l =\sqrt{2/eB} =1$)
and the first  factor is the  correlation function of the electron hole field operators computed in the $\Z_3$ parafermion CFT.
 $N$ is the number of electrons with coordinates $z_i$  and must be equal to $3r$ with $r$ an integer because the electron fields  
have a non-trivial  $\Z_3$ parafermion charge and must be created in triples \cite{rr,NPB2001}. 
Here $\psi_{\mathrm{hole}}(z_i)$ is the 
electron hole operator \cite{NPB2001} representing an electron hole with coordinate $z$
\beq\label{psi_h}
\psi_{\mathrm{hole}}(z)=\np{\exp\left(i\sqrt{\frac{5}{3}}\phi(z)\right)} \ \psi_1(z), \quad \psi_1(z) \equiv \Phi(\L_1+\L_1)(z)
\eeq
of a $\uu$ vertex exponent of a normalized chiral boson field  with charge $1/\sqrt{\nu_H}$  and the $\Z_3$ parafermion primary field 
$\psi_1(z) \equiv \Phi(\L_1+\L_1)(z)$ whose CFT dimension is $2/3$ in the notation of \cite{NPB2001}. 
The electric charge of the electron field is $1$ (in units in which $e=h =1$) and because of the charge--flux relation (\ref{charge-flux})
it carries fractional magnetic flux - the quantization of the magnetic flux in FQH liquids is the physical reason why the electrons in the ground state 
must be created in triples.
The total conformal dimension of the electron is 
\[
\Delta_{\mathrm{hole}}=\frac{1}{2}\left(\sqrt{\frac{5}{3}}\right)^2 + \frac{2}{3}=\frac{3}{2}
\]
 as it should be for a fermion field. It is worth emphasizing that the CFT dimension of the $\uu$ part of the electron operator is fixed by 
the filling factor $\nu_H=3/5$ to be $(1/2)(5/3)$
 and twice this number is the statistical angle of the electron field. It is obvious that there should be a non-trivial 
neutral component of the electron field operator so that the total CFT dimension would be half-odd integer. 
This neutral part of the electron field here is given by the parafermion current $\psi_1(z)$.
 
 The state denoted by $\la Q_{\mathrm{bg}} |$ is the CFT conjugate  \cite{CFT-book} of the CFT state in the  occupation number representation 
which can be obtained by fusing all charged and fields to the center $z_0=0$ of the FQH disk
 \[
 | Q_{\mathrm{bg}} \ra  = \lim_{z_0\to 0} \np{\exp\left(i  Q_{\mathrm{bg}} \phi(z_0)\right)} )|0\ra
 \]
 where $Q_{\mathrm{bg}}=  N\sqrt{\frac{5}{3}}$ is the $\uu$ charge of the state with all electric charges 
concentrated at the origin of the disk.
Using the CFT rule for state conjugation  \cite{CFT-book} we have
 \[
 \la Q_{\mathrm{bg}} | = \lim_{z_0\to \infty} z_0^{2\Delta_l }
 \la 0 | \np{\exp\left(-i  Q_{\mathrm{bg}} \phi(z_0)\right)}  
 \]
 with $\Delta_l = (Q_{\mathrm{bg}})^2/2 =5N^2/6$.

 Traditionally in the literature on QH states the quantity $Q_{\mathrm{bg}}$ has been called the background charge or screening charge
 which is plugged into the CFT correlation function in order to  make it non-zero. However $Q_{\mathrm{bg}}$ in our case plays
a more fundamental role because it can be interpreted as the  representation of the state of the FQH system in the occupation number representation. 
Because we are interested in the braid behavior  of the anyons we need a coordinate many-body wave function  $\Phi(x_1,\ldots, x_N)$  
which can be obtained from the quantum state $|\Phi_N\ra = |Q_{\mathrm{bg}}\ra $ in the occupation number representation 
using the standard relation  (cf. Eq.~(117) in Ref.~\cite{schweber})
\beqa\label{Phi_N}
&&|\Phi_N\ra=\frac{1}{\sqrt{N!}} \int \dd {\bf x}_1 \ldots \dd {\bf x}_N
\   \Phi({\bf x}_1,\ldots, {\bf x}_N)  \ \widehat{\Psi}^\dagger({\bf x}_N) \ldots \widehat{\Psi}^\dagger({\bf x}_1)|0\ra
\Longleftrightarrow \nn
&& \Phi^*({\bf x}_1,\ldots, {\bf x}_N) = \frac{1}{\sqrt{N!}}
\la \Phi_N|  \widehat{\Psi}^\dagger({\bf x}_N) \ldots \widehat{ \Psi}^\dagger({\bf x}_1)|0\ra,
\eeqa
where  $\widehat{ \Psi} ({\bf x})=\sum_{n}\hat{a}_n  \psi_n({\bf x})$ is the electron field operator ($\hat{a}_n, n=0,\ldots, \infty$ are the 
fermionic annihilation operators) satisfying  together with its Hermitian conjugate  
$\widehat{\Psi}^\dagger({\bf x})$ the 
canonical anticommutation relations $\left\{ \widehat{\Psi}({\bf x}), \widehat{\Psi}^\dagger ({\bf x}) \right\}= \delta ({\bf x}-{\bf x}')$ \cite{schweber}.
 
 
In a similar way we shall construct the coordinate wavefunction for $N$ electron holes and $4$ Fibonacci anyons, with coordinates $w_a$ in the plane, 
which we shall use as qubit (from now on we skip the standard Gaussian exponent and leave only the holomorphic part)
\beq\label{Phi_4F}
\Phi_{4F}(\{w_a\}; \{z_i\})=\la Q_{\mathrm{bg}},\Lambda | \psi_F(w_1)\psi_F(w_2)\psi_F(w_3)\psi_F(w_4) \prod_{i=1}^{N} \psi_{\mathrm{hole}}(z_i) | 0 \ra,
\eeq
where $\psi_F(w)$ is the field operator representing a Fibonacci anyon with coordinate $w$ in the complex plane as a primary field in 
the $\uu \times \PF_3$ CFT
\beq\label{field-eps}
\psi_F(w)=   \np{\e^{i\frac{3}{\sqrt{15}} \phi(w)}} \   \varepsilon(w)
\eeq
with  $\varepsilon(w)$ being the coset CFT primary field of dimension $\Delta=2/5$, labeled by the weight $\L_1+\L_2$,
 with coordinate $w$ in the complex plane. We point out here that the $\Z_3$ parafermion charge of the field  $\varepsilon(w)$
 is $0$ and the magnetic flux corresponding to $\varepsilon(w)$ is integer. Therefore there is no clustering of  the $\varepsilon(w)$ fields and 
all correlation functions of arbitrary number of  $\varepsilon(w)$ might be non-zero (again the number of electrons 
 must be $N=3r$ with $r$ integer).
 It is important to note also that the Fibonacci anyon field (\ref{field-eps}) must be relatively
 local with the electron field (\ref{psi_h}), see assumption (A4) in  \cite{fro-stu-thi}, which is indeed the case as can be seen from the 
operator product expansion (OPE normalization is the same as in Ref.~\cite{ardonne-schoutens})
 \beq\label{locality}
 \psi_{\mathrm{hole}}(z)\psi_F(w) \mathop{\simeq}_{z\to w} \frac{2}{\sqrt{3}}\  \np{\e^{i\frac{8}{\sqrt{15}} \phi(w)}} \ \sigma_2(w) +O(z-w)
 \eeq
  The occupation number representation state denoted by $\la Q_{\mathrm{bg}},\Lambda |$ is now the CFT conjugate of the CFT state which is this time 
obtained by fusing all charged and parafermion fields to the origin $z_0=0$
 \[
 | Q_{\mathrm{bg}},\Lambda \ra  = \lim_{z_0\to 0} \np{\exp\left(i  Q_{\mathrm{bg}} \phi(z_0)\right)} \ \Phi_\Lambda(z_0)|0\ra
 \]
 where $Q_{\mathrm{bg}}= 4\frac{3}{\sqrt{15}} +N\sqrt{\frac{5}{3}}$ is the $\uu$ charge of the state with all electric charges concentrated at the origin (of the FQH disk) and 
 $\Phi_\Lambda(z_0)$ is the $\Z_3$ parafermion field with weight $\Lambda$ which is the result of  the fusion of all parafermion topological charges at the origin. 
Using the CFT rule for state conjugation \cite{CFT-book} we have
 \[
 \la Q_{\mathrm{bg}},\Lambda | = \lim_{z_0\to \infty} z_0^{2(\Delta_l +\Delta_\Lambda)}
 \la 0 | \np{\exp\left(-i  Q_{\mathrm{bg}} \phi(z_0)\right)} \ \Phi^*_{\Lambda}(z_0) \ 
 \]
 with $\Delta_l = (Q_{\mathrm{bg}})^2/2 =(12+5N)^2/30$ and $\Delta_\Lambda$ is the CFT dimension of the parafermion field $\Phi_\Lambda(z_0)$.
 
 Writing explicitly the correlation function of the $\uu$ part, which is of Laughlin type, and separately the correlation function in 
the $\Z_3$ parafermion part,
we obtain the wave function for 4 $\varepsilon$ fields and $N=3r$ electrons with $r$ integer (we denote by $\la \Lambda |$ the 
CFT conjugate of  $ | \Lambda\ra = \Phi_\Lambda(0)|0\ra$ )
 \beqa \label{decomposed}
\Psi_{4 F}(w_1,\ldots, w_4; z_1,\ldots, z_N)=
\la \Lambda|  \varepsilon(w_1) \varepsilon(w_2)\varepsilon(w_3)\varepsilon(w_4) \prod_{i=1}^{N} \psi_{1}(z_i) |0 \ra_{\PF}  \times \nn
\prod_{1 \le a< b \le 4 } (w_a-w_b)^{\frac{3}{5}}
 \prod_{a=1 }^{4}\prod_{i=1 }^{N} (w_a-z_i) \prod_{1 \le i<j \le N} (z_i-z_j)^{\frac{5}{3}} \qquad
\eeqa
 where now the remaining correlation function is only in  the (neutral) parafermion CFT and again $\psi_1(z)$ is given in Eq.~(\ref{psi_h}).
 Following Ref.~\cite{ardonne-schoutens} and using this decomposition of the 4 Fibonacci anyon wave function we will obtain the 
braid matrices for exchanging Fibonacci anyons in Sect.~\ref{sec:braid}.

As can be seen from  Fig.~\ref{fig:brattelli-fibonacci} the register containing $n$ Fibonacci anyons $\varepsilon$ can fuse at the 
origin of the FQH disk either to $\I$ or to $\varepsilon$
 so that the weight in Eq.~(\ref{decomposed}) can only be $\Lambda = \I$, or  $\Lambda = \varepsilon$. This is an overall 
topological characteristics of the quantum register which can be interpreted as a $\Z_3$ topological charge at infinity.
It  can be measured in electronic Fabri--Perot interferometers by transporting single 
non-Abelian quasiparticles ($\sigma_1$ or $\sigma_2$) around the register, see Sect.~\ref{sec:FPI}.
This topological characteristics cannot be changed during the processes of braiding. Therefore for any number $n$ of $\varepsilon$ fields there will be two decoupled  
representations of the braid group $\B^{(n)}$ which we shall label by a subscript of the left bra-vector. For $n=4$ we have
\beq\label{top-charge}
{}_{\I}\la \varepsilon(w_1) \varepsilon(w_2)\varepsilon(w_3)\varepsilon(w_4)  \ra, \quad
{}_{\varepsilon}\la \varepsilon(w_1) \varepsilon(w_2)\varepsilon(w_3)\varepsilon(w_4)\ra
\eeq
 where we have skipped the product $\prod_{i=1}^{N} \psi_{1}(z_i)$ for convenience.

There is one simple but very important observation that correlation function of a single $\varepsilon$ field and an arbitrary product of 
parafermion fields  $\psi_{1}(z_i)$ vanishes
\[
{}_{\I} \left\la \varepsilon(w) \prod_{i=1}^{N} \psi_{1}(z_i) \right\ra_{\PF}  \equiv 0 ,  \quad
{}_{\varepsilon} \left\la \varepsilon(w) \prod_{i=1}^{N} \psi_{1}(z_i) \right\ra_{\PF}  \neq 0 
\]
while the correlation function of a single $\varepsilon$ field and an arbitrary product of 
parafermion fields  $\psi_{1}(z_i)$ in the $\varepsilon$-representation can be non-zero.
Because of the $\Lambda=\varepsilon$ case in Eq.~(\ref{top-charge}) a single $\varepsilon$ field can indeed be localized on an 
antidot which is the initialization scenario for the Fibonacci quantum computer, see Sect.~\ref{sec:ini}.

From now on we will consider only the  representations corresponding to $\Lambda=\I$. 
Using the relative locality  of the fundamental $\Z_3$ quasihole $\psi_{\mathrm{qh}}(w)$  with the electron field  (\ref{psi_h}), namely
\beq\label{relative-qh}
\psi_{\mathrm{qh}}(w)=   \np{\e^{i\frac{1}{\sqrt{15}} \phi(w)}} \   \sigma_1(w), \quad
 \psi_{\mathrm{hole}}(z)\psi_{\mathrm{qh}}(w) \mathop{\simeq}_{z\to w} \sqrt{\frac{2}{3}} \  \np{\e^{i\frac{6}{\sqrt{15}} \phi(w)}} \ \varepsilon(w)
\eeq
we can express the $\Z_3$ parafermion CFT 
blocks $\Psi_{4\varepsilon}(w_a; z_i)$ corresponding to the basis vectors, labeled by superscript $0$ or $1$, of the single-qubit 
space (\ref{decomposed}) in terms of the $\Z_3$ parafermion correlation function of $4$ fields $\sigma_1$ and $3r+4$ fields $\psi_1$ as follows
\beqa \label{Psi_01}
\Psi_{4\varepsilon}(w_a; z_i)^{(0,1)}= \la \varepsilon(w_1)\varepsilon(w_2)\varepsilon(w_3)\varepsilon(w_4) 
\prod_{i=1}^{3r} \psi_1(z_i) \ra_{\PF}^{(0,1)} = \nn
\frac{9}{4} \prod_{1\leq a < b \leq 4} (w_a-w_b)^{-\frac{1}{3}} \prod_{a=1}^4 \prod_{i=1}^{3r} (w_a-z_i)^{-\frac{5}{3}} \times \nn
\lim_{w'_a \to w_a}\la \sigma_1(w_1)  \sigma_1(w_2) \sigma_1(w_3) \sigma_1(w_4) \prod_{a=1}^4 \psi_1(w'_a) 
\prod_{i=1}^{3r} \psi_1(z_i) \ra^{(0,1)}_\PF .
\eeqa
Following \cite{NPB2001}, the last correlation function of 4 quasiholes and $N = 3 r+ 4$ electrons in Eq.~(\ref{Psi_01}) was expanded in 
Ref.~\cite{ardonne-schoutens} in terms of the functions $\Psi_{(12)(34)} (w_a , z_a, z_i)\,$ and $\Psi_{(13)(24)} (w_a , z_a,  z_i)\,$ reflecting the two 
independent ways of splitting the 4 anyons into two pairs (like in the Moore--Read FQH state \cite{mr}). The coefficient functions 
(depending on $w_a\,, a= 1,\dots , 4$) computed in \cite{ardonne-schoutens} were expressed in terms of hypergeometric functions \cite{bateman-erdelyi} of 
certain harmonic ratio. 

These results were used in Ref.~\cite{HG-Monodromy} to present, after taking the limit $w'_a \to w_a\,,$ the correlation function (\ref{Psi_01}) for any integer $r\,$ as
\beq\label{4-eps-01}
\la \varepsilon(w_1)\varepsilon(w_2)\varepsilon(w_3)\varepsilon(w_4) 
\prod_{i=1}^{3r} \psi_1(z_i) \ra_{\PF}^{(p)} = \Phi^{(p)}(\{w\},\{z\})\ ,\quad p=0,1
\eeq
where  
\beqa
\Phi^{(0)}&&(\{w\} , \{ z \} ) = Q \,[(w_{12} w_{34})^{-3}\,\eta^3\, (1-\eta)^{-\frac{3}{2}} ]\, \left( w_{12} w_{34}\right)^{- \frac{4}{5}} (1-\eta)^{\frac{1}{10}}\,\times\nn
&&\times\, \Big[\, F\Big(\frac{1}{5} , \frac{4}{5} , \frac{3}{5} ; \eta \Big)\, \Psi_{12,34} -\frac{1}{3} \,F\Big(\frac{6}{5} , \frac{4}{5} , \frac{8}{5} ; \eta \Big)\, \Psi_{13,24}\, \Big] \ ,\qquad\ \label{4e-eta0}\\
\Phi^{(1)}&&(\{w\} , \{ z \})  = C\,Q\,[(w_{12} w_{34})^{-3}\,\eta^3\, (1-\eta)^{-\frac{3}{2}} ]\left( w_{12} w_{34}\right)^{- \frac{4}{5}} \eta^{-\frac{3}{5}} (1-\eta)^{- \frac{3}{10}}\,\nn
&&\times\, \Big[\,\eta\, F\Big(\frac{1}{5} , \frac{4}{5} , \frac{7}{5} ; \eta \Big)\,\Psi_{12,34} - 2 \,F\Big(\frac{1}{5} , -\frac{1}{5} , \frac{2}{5} ; \eta\Big )\, \Psi_{13,24}\, \Big]\ .
\label{4e-eta1}
\eeqa
Here $w_{ab}=w_a-w_b$,  $\eta=\frac{w_{12}w_{34}}{w_{13}w_{24}}$,  $2C= \sqrt{\frac{\Gamma(1/5)\Gamma^3(3/5)}{\Gamma(4/5)\Gamma^3(2/5)}}$ and
\beq
Q = Q(\{w\}, \{ z\}) := \frac{9}{4} \,\prod_{a=1}^4 \prod_{i=1}^{3r} (w_a-z_i)^{-1} \,\prod_{1\le j < k \le 3r} (z_j - z_k)^{- \frac{2}{3}}\ ,
\label{Qwuz}
\eeq
with $\{w\} = \{ w_1, \dots , w_4\}$ and $\{ z\} = \{ z_1 , \dots , z_{3r} \}\,.$ The factors $\Psi_{12,34}\,$ and $\Psi_{13,24}\,$ are polynomials in 
(the differences of) $\{w \}$ and $\{z \}\,$ defined as
\beqa
&&\Psi_{12,34} = \Psi_{12,34}(\{w\},\{z\}) := 
\Psi_{(12)(34)} (w_a, w_a, z_i)\ , \nn
&&\Psi_{13,24} = \Psi_{13,24}(\{w\},\{z\}) := 
\Psi_{(13)(24)} (w_a, w_a, z_i) \ .
\label{Psi_1234}
\eeqa
The result in Eq.~(\ref{4-eps-01}) was obtained (for slightly different limits) in Ref.~\cite{ardonne-schoutens} only for $r=0\,.$ However, a wave function 
representation in terms of CFT correlators like the one in Eq.~(\ref{Phi_4F}) only has physical meaning describing the FQH state in the thermodynamic 
limit $N = 3\,r \to \infty\,.$ The needed generalization to any positive integer $r\,$ (presumably, $r\gg 1$) has been carefully derived in Ref.~\cite{HG-Monodromy}. 

The 4-anyon wavefunctions (\ref{4e-eta0}) and (\ref{4e-eta1}) will be our starting point for the derivation of the generators of the braid group ${\cal B}_4\,$ for 4 
Fibonacci anyons in the $\Z_3\,$ parafermion FQH state.
To this end, one does not need to know the explicit form of the factors (\ref{Psi_1234}). It is sufficient to take into account the Nayak-Wilczek type relation 
connecting the ones corresponding to the three possible ways of splitting 4 anyons into two pairs which, written in terms of the harmonic ratio $\eta$ reads
\beq
\eta\, \Psi_{12,34} - \Psi_{13,24} + (1-\eta)\,\Psi_{14,23} = 0
\label{NWrel}
\eeq
with $\Psi_{14,23}\,$ defined analogously to (\ref{Psi_1234}), cf. Refs.~\cite{ardonne-schoutens,HG-Monodromy}.

\section{Fibonacci anyons in the $\Z_3$ parafermion  FQH state}
\label{sec:fibonacci-Z_3}
\subsection{Initialization of the Fibonacci qubit}
\label{sec:ini}
\begin{figure}[htb]
	\centering
	\includegraphics[viewport=70 330 485 520,clip,width=\textwidth]{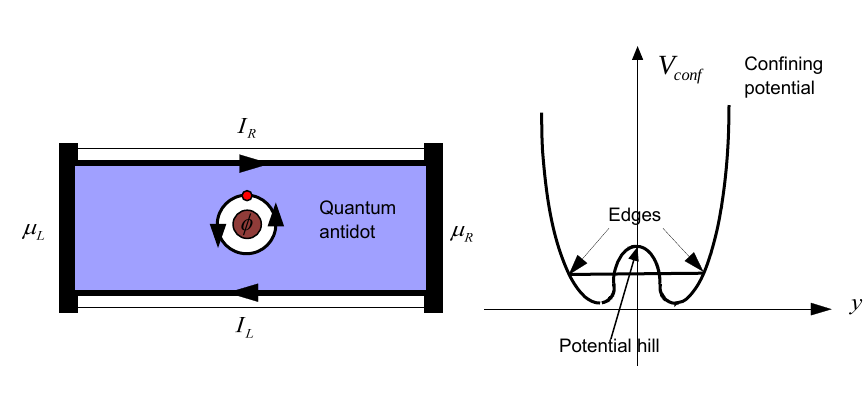} 
\caption{An antidot is an island (in white)  inside of a quantum Hall liquid (in blue) with filling factor $\nu_H=3/5$.
The arrows show the direction of the edge currents along the edges and along the perimeter of the antidot. 
The chemical potentials on the left and right contacts are denoted by $\mu_L$ and $\mu_R$.
 \label{fig:antidot}}
\end{figure}

The construction of the qubit for Fibonacci anyons is similar to that for the Ising anyons proposed by 
Das Sarma et al. \cite{sarma-freedman-nayak}. 
Consider a quantum antidot inside of a FQH bar with $\Z_3$ parafermion liquid with $\nu_H=3/5$, as shown in Fig.~\ref{fig:antidot}.
The antidot is created by adding a potential hill, shown on the right  in  Fig.~\ref{fig:antidot}, to the confining potential $V_{conf}$ 
creating the FQH bar. This potential hill creates an island in the FQH liquid whose edge accommodates edge current due to the 
magnetic field.
Next, we increase  adiabatically the magnetic flux threading the antidot until it reaches one quantum of Aharonov--Bohm  flux $\phi=h/e=1$ 
 where $h$ is the Plank constant and $e$ is the charge of the electron (we can think of the flux as created by an  infinitely thin solenoid). 
This is depicted in Fig.~\ref{fig:antidot} by the brown disk inside the antidot with the symbol $\phi$ inside it.
Now the incompressible quantum Hall liquid surrounding the antidot responds to the flux threading the antidot by localizing one or 
several quasiholes/quasiparticles along the edge of the antidot, depicted by the small red circle on the antidot's edge, 
in such a way to compensate the flux $\phi$. The electric charge of the 
quasiparticle localized on the antidot is uniquely determined by the universal quantum Hall flux--charge relation \cite{fro-stu-thi,NPB2001}
\beq\label{charge-flux}
Q_\el = \nu_H \ \phi = \frac{3}{5}
\eeq
in units of the electron charge $e$. The most elegant way to see why this happens is by using the Laughlin 
argument about adding one flux quantum \cite{laughlin} corresponding to the spectral flow \cite{cz}.
The magnetic flux $\phi$ threading the antidot is quantized in units of $h/e$ and the adiabatic increase of $\phi$ from $0$ to $1$ transfers 
 electric charge equal to the Hall filling factor $\nu_H$ from the outer edge to the antidot's edge \cite{cz}. 
There are only two quasiparticle excitations in the spectrum of the $\Z_3$ parafermion FQH liquid with electric charge $3/5$. 
If we denote the excitation's label in the full CFT
for the $\Z_3$ parafermionic states by the weight
\beq\label{lambda}
\underline{\l} = l \underline{\mathrm{e}}^*_1 + \L_\mu+\L_\nu, \quad 0\leq \mu \leq \nu \leq 2, \quad \mathrm{where} \quad l=\mu+\nu \mod 3
\eeq
in the notation of Ref.~\cite{NPB2001}, then the electric charge of this excitation is equal to $Q_\el(\underline{\l}) = l/(kM+2) = l/5$ for $k=3$
and $M=1$. Therefore the excitations with which the FQH liquid could respond to the flux threading must have $l=3$. There are only two such excitations which
can be characterized by the full RCFT characters, or chiral partition functions (\ref{full-ch}). As can be seen from Eq.~(A.1) in  Appendix A of Ref.~\cite{NPB-PF_k} 
the first one is characterized by the chiral partition function 
\beq\label{ch_I}
 K_{3}(\t,3\z;15) \ch(\L_0+\L_0)(\t), 
\eeq
 i.e., $l=3$,  $\mu=\nu =0$ in Eq.~(\ref{lambda}) above  and corresponds to the chiral vertex operator
\beq\label{field-I}
\np{\e^{i\frac{3}{\sqrt{15}} \phi(z)}} \ \Phi(\L_0+\L_0), \quad \mathrm{where} \quad \Phi(\L_0+\L_0) = \I
\eeq
is the parafermion field corresponding to the $\Z_3$ parafermion vacuum and the chiral boson $\phi(z)$ in the $\uu$
vertex exponent, representing the electric charge part of the RCFT, is normalized by (\ref{normal}).
The second excitation which can be localized on the antidot's edge, in response to the flux threading, is characterized by the 
chiral partition function \cite{NPB2001,NPB-PF_k}
\beq\label{ch_eps}
K_{3}(\t,3\z;15) \ch(\L_1+\L_2)(\t) 
\eeq
 i.e., $l=3$,  $\mu=1$, $\nu =2$ in Eq.~(\ref{lambda}) and corresponds to the chiral vertex operator (\ref{field-eps}) where $\Phi(\L_1+\L_2) = \varepsilon$
is the parafermion field corresponding to the Fibonacci anyon. It is obvious from the explicit list of characters in  Eq.~(A.1) in  
Appendix A of Ref.~\cite{NPB-PF_k} that there are no other
$l$, $\mu$ and $\nu$ satisfying the pairing rule given in the right-hand side of Eq.~(\ref{lambda}).

\textbf{Remark:}
The $\Z_3$ pairing rule (\ref{PR}) repeated in (\ref{lambda}) in the diagonal coset model is naturally inherited from its Abelian parent.
In the framework of the so called Maximally Symmetric Chiral Quantum Hall Lattices, see \cite{NPB2001} and references therein, 
the topological charges of the multielectron-like excitations form an odd integer lattice. The structure of this lattice is almost completely determined by the
filling factor $\nu$, the maximal symmetry and the locality conditions. The elements in the dual lattice represent the topological charges of quasiparticle excitations.
However, in order to separate the $\uu$ charges from the neutral degrees of freedom we need to have a direct sum decomposition. The almost unique 
lattice $\Gamma$ which is derived in \cite{NPB2001}, for the filling factor $\nu=k/(k+2)$ (in our case $k=3$), is not completely decomposable. 
Nevertheless, there is a decomposable sublattice $L\subset \Gamma$ in which the one-dimensional charge
lattice is completely separated from the neutral sublattice. Because of the inclusion $L\subset \Gamma \subset \Gamma^* \subset L^*$, the 
topological charges of the various super-selection sectors of the original lattice, which belong to the finite Abelian group $\Gamma^*/\Gamma$,
form a subgroup of the decomposable group $L^*/L$, which is bigger than the physical one. 
The pairing rule Eq.~(3.16) in \cite{NPB2001} selects those charges of $L^*/L$ which belong to the 
original physical super-selection sectors of $\Gamma^*/\Gamma \subset L^*/L$, see . The pairing rule (\ref{PR}) is the descendant of the pairing rule in the 
Abelian parent after the coset projection is done. That is why the diagonal coset approach to the $\Z_3$ parafermions is more appropriate for the 
Fibonacci anyons initialization, like in (\ref{ch_eps}),  than the others.

It is worth emphasizing here that the chiral partition functions (\ref{ch_I}) and (\ref{ch_eps}) are not sums over the entire
orbit of the simple current (\ref{psi_h}) like Eq.~(\ref{full-ch})  because we assume that the antidot is well separated from the outer edge or 
contacts so that electron tunneling into the edge of the antidot is not expected. If there is electron tunneling into the antidot then the full 
chiral partition functions would be $\chi_{-2,2}$ instead of Eq.~(\ref{ch_I}) and $\chi_{-2,1}$ instead of Eq.~(\ref{ch_eps}) in the notation of 
Eq.~(\ref{full-ch}).
This sum in Eq.~(\ref{full-ch}) represents the simple physical observation that,
in the thermodynamic limit in which the number of electrons goes to infinity,
adding one electron/hole to the edge of the chiral QH sample (a disk),
does not change anything in the system and should 
be a symmetry of all thermodynamic quantities computed in the effective CFT. 

The modular $S$ matrix has well-known transformation properties under the action of the simple currents \cite{CFT-book} 
and as a result the entire orbit under the action of the simple currents has the same $S$-matrix elements \cite{JGSP-S-PF}. 
Notice also that the quantum dimensions of all parafermion fields in
Eq.~(\ref{ch_I}) are $1$,  and the fields $\Phi(\L_0+\L_0)= \I$,  $\Phi(\L_1+\L_1)=\psi_1$ and $\Phi(\L_2+\L_2)=\psi_2$ 
are Abelian (the parafermion currents), while the parafermion fields  $\Phi(\L_0+\L_1)= \s_1$,  $\Phi(\L_0+\L_2)=\s_2$ 
and $\Phi(\L_1+\L_2)=\varepsilon$  in Eq.~(\ref{ch_eps}) all have quantum dimension $\delta$ and are non-Abelian.
This is the physical explanation of the ``coarse-graining''  of the fusion rules formulated  in Ref.~\cite{ardonne-schoutens}.

The process of localization of a quasiparticle represented by the field operators (\ref{field-I}) or (\ref{field-eps}) on the
antidot's edge, in response to piercing the antidot by one quantum of magnetic flux, is stochastic. In some cases the 
localized quasiparticle from the parafermion point of view would be $\I$, in the others it would be $\varepsilon$ and no other 
possibilities exist because of the $\Z_3$ paring rule \cite{NPB2001}. Following the alternative interpretation of the quantum 
dimensions \cite{preskill-TQC}, according to which the probability to create anyons of type $a$ and quantum dimension $d_a$
in a random process is (see Eq.~(9.112) in Ref.~\cite{preskill-TQC})
\[
p_a = \frac{d_a^2}{D^2}, \quad \mathrm{where} \quad D^2 = \sum_a d_a^2
\]
 is the total quantum dimension, we expect that the probability to create a Fibonacci anyon $\varepsilon$ is bigger than that to
 create the unit $\I$  because $p_{\varepsilon}/ p_{\I} = \delta^2=\delta+1 \approx 2,618$.
 
 The topological qubit for quantum computation based on Fibonacci anyons, constructed in the basis (\ref{4F-comp-basis}), 
can be implemented with 4 antidots, each of which is 
threaded  with one quantum of magnetic flux and in which all quasiparticles localized on the antidots are $\varepsilon$. This can be 
achieved by measuring non-destructively each antidot, as shown in the next Subsection, and if on some of the antidots the edge 
quasiparticle is $\I$ the stochastic process of initialization is repeated until an $\varepsilon$ excitation is registered. 

Notice that this initialization of the quantum state with a Fibonacci anyon is simpler than the initialization of the Ising anyon 
state \cite{sarma-freedman-nayak} because in the latter the insertion of one quantum of magnetic flux creates two types of 
excitations which cannot form coherent superpositions due to their opposite fermionic parity. The problem there is solved if we start
with two antidots, then insert one flux quantum through one of the antidot and finally apply small voltage between the two antidots
to induce tunneling of a single Ising anyon between the two antidots  \cite{sarma-freedman-nayak,TQC-NPB}. Here the problem 
of initialization of the Fibonacci anyon is solved much simpler and easier by selecting only those antidots which contain an $\varepsilon$ field
or, by repeating the process of flux insertion until an $\varepsilon$ field is obtained, i.e., no tunneling is needed.
\subsection{Measurement of the Fibonacci anyon}
\label{sec:FPI}
\begin{figure}[htb]
	\centering
	\includegraphics[viewport=20 210 770 630,clip,width=13cm]{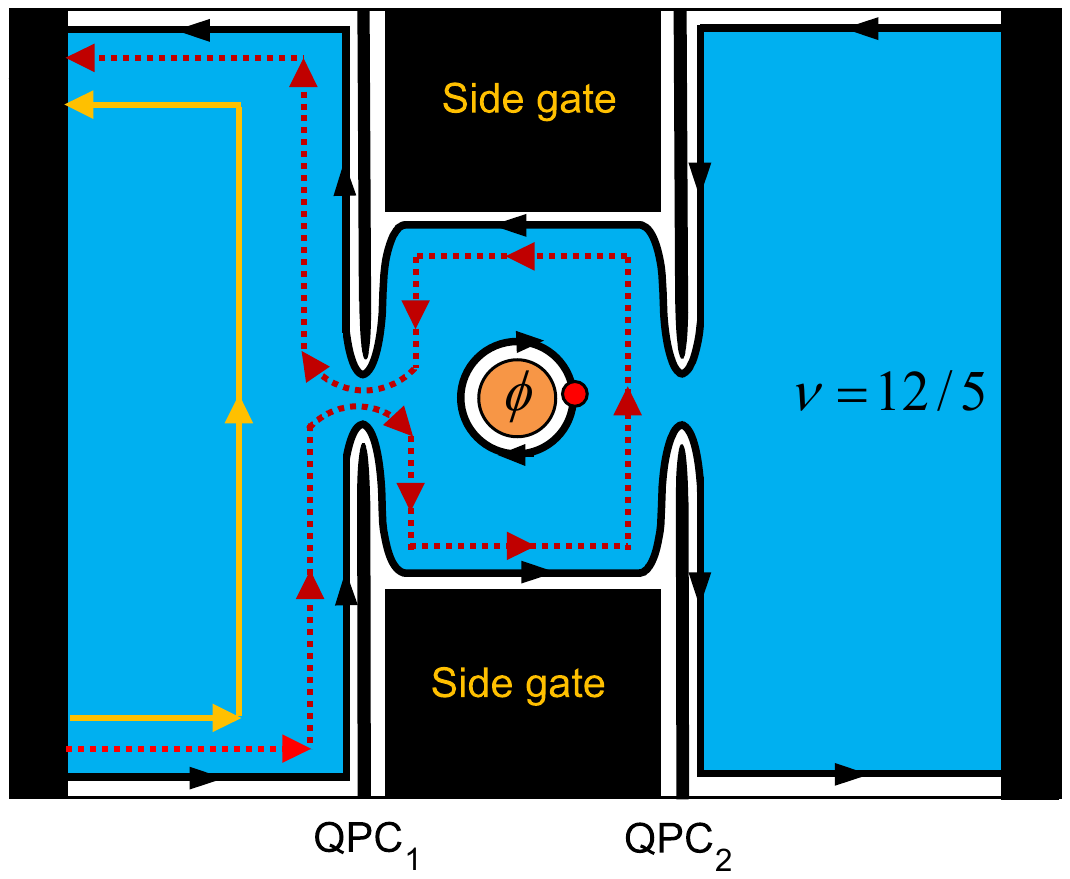} 
\caption{Measurement of an Fibonacci anyon by Fabri-Perot interferometer
 \label{fig:FPI-Fibonacci}}
\end{figure}

%
One possibility to detect  the Fibonacci anyon on the antidot is to look to the interference patterns in Fabry--P\'erot interferometers 
in the backscattered current shown in Fig.~\ref{fig:FPI-Fibonacci}.
Here the two quantum point contacts (QPC), denoted as   QPC$_1$ and QPC$_2$,  are not completely pinched off and there is a
small current
reflected back from the interferometer, due to the tunneling of quasiparticles through the QPCs, while almost all electric charge is
transmitted through the interferometer along the edge states denoted by arrows in Fig.~\ref{fig:FPI-Fibonacci}.
To lowest order in the amplitudes $t_1$ and $t_2$, for tunneling of fundamental quasiparticles through quantum-point contacts
QPC$_1$ and QPC$_2$ respectively, the  amplitude of the backscattered current of the Fabry--P\'erot  interferometer shown
 in Fig.~\ref{fig:FPI-Fibonacci}, in the quantum state $|\Psi\ra$ of the strongly correlated  FQH electron system, is proportional
to the ``diagonal'' conductivity \cite{bonderson-5-2,JGSP-S-PF}
\beqa \label{s_xx}
\s_{xx}  & \propto & || \left( t_1 U_1 + t_2 U_2 \right) |\Psi\ra ||^2 =
\la \Psi |   (t_1^* U_1^\dagger + t_2^* U_2^\dagger)\left( t_1 U_1 + t_2 U_2\right) |\Psi\ra \nn
& = & |t_1|^2 + |t_2|^2
 +2 \mathrm{Re} \left( t_1^* t_2 \la \Psi |   U_1^{-1} U_2 |\Psi\ra \right).
\eeqa
The matrix element appearing in Eq.~(\ref{s_xx}) of the two unitary operators $U_1$ and $U_2$, each of which represents
the quasiparticle evolution in the  state $|\Psi\ra$  during the process of tunneling
through QPC$_1$ and  QPC$_2$ respectively,  determines the interference effects and can be written as \cite{bonderson-5-2}
\beq\label{alpha}
 \la \Psi |   U_1^{-1} U_2 |\Psi\ra = \e^{i\alpha}  \la \Psi |  ( B_1)^{2}|\Psi\ra  = \e^{i\alpha}  \la \Psi |   M |\Psi\ra 
\eeq
where $\alpha$ is an Abelian phase which is a sum of the dynamical phase associated with the unitary evolution of the quasiparticle
 transported
along the full path around the central region (the island) of the interferometer containing $n$ fundamental quasiparticles and  the
topological phase due to the Aharonov--Bohm effect of the electrically charged quasiparticles in the total magnetic field. The expectation
value of $(B_1)^2\equiv M$  represents only the action of the pure braiding operator taking the traveling quasiparticle around the
static quasiparticles localized  in the central region. While for Abelian quasiparticle of type $a$ transported along a complete loop around
quasiparticle of type $b$ this monodromy always
satisfies  $| \la \Psi |   M_{ab} |\Psi\ra|=1$, the monodromy expectation value  for non-Abelian anyons  $| \la \Psi |   M_{ab} |\Psi\ra|\leq 1$
and could eventually be 0, which corresponds to no interference at all \cite{stern-halperin-5-2,bonderson-5-2}.

To the lowest order in the tunneling interference process, the monodromy expectation value for a
quantum state $|\Psi_{ab}\ra$ of uncorrelated quasiparticles of type $a$ and $b$, can be computed exactly \cite{bonderson-12-5}  in
terms of the  modular $S$ matrix \cite{CFT-book} according to
\beq \label{M}
 \la \Psi_{ab} |   M |\Psi_{ab}\ra = \frac{S_{ab} S_{00}}{S_{0 a} S_{0b}}
\eeq
where $S_{ab}$ is the matrix element of the modular $S$ matrix corresponding to the topological charges $a$ of the quasiparticle being
transported along a complete loop around a quasiparticle of topological charge $b$, while $0$ labels the vacuum sector,
i.e., the state without any quasiparticle. Therefore if we know the $S$ matrix explicitly we can compare all interference patterns
 corresponding
to given types of static quasiparticles localized in the central region of the interferometer trying in this way to extract information about
monodromy matrix elements and proving or disproving the emergence of non-Abelian quasiparticles in each experimental setup.
We can use the explicit form of the modular $S$ matrix for the diagonal coset $\PF_3$ derived in \cite{JGSP-S-PF}, see Eq.~(36) there,
to calculate the interference terms in the diagonal conductivity.
When we apply small voltage between the two QPCs the most relevant quasiparticles $\sigma_1$ or $\sigma_2$ will tunnel creating in this 
way a monodromy transformation around the antidot. If  the localized quasiparticle on the  antidot is the field (\ref{field-I}) then the parafermion
 (or neutral part of the) monodromy expectation value will be 
\[
 \la \Psi_{04} |   M |\Psi_{04}\ra = \frac{S_{04} S_{00}}{S_{0 0} S_{04}}=1
\]
because the parafermion primary field $\Phi(\L_0+\L_1)=\sigma_1$ (the 4-th basis vector of the $S$ matrix) is transported along a closed loop
around the  parafermion primary field $\Phi(\L_0+\L_0)=\I$ (the 4-th basis vector of the $S$ matrix).
If on the other hand the localized quasiparticle on the  antidot is the field (\ref{field-eps}) then the parafermion
 (or neutral part of the) monodromy expectation value will be 
\[
 \la \Psi_{64} |   M |\Psi_{64}\ra = \frac{S_{64} S_{00}}{S_{0 6} S_{04}}=-\frac{1}{\delta^2}
\]
because the parafermion primary field $\Phi(\L_0+\L_1)=\sigma_1$ (the 4-th basis vector of the $S$ matrix) is transported along a closed loop
around the  parafermion primary field $\Phi(\L_1+\L_2)=\varepsilon$ (the 6-th basis vector of the $S$ matrix).
Therefore if the Anyon localized on the antidot is Abelian - there is no suppression of the interference term; if the anyon is non-Abelian - there is a 
suppression of the interference term by a factor of  $1/\delta^2\approx 0,38$. The same result is obtained if the tunneling quasiparticle between the 
QPCs is $\sigma_2$.

Notice that this measurement does not change or destroy the state of the anyon on the antidot.
So with every antidot we repeat the initialization step of adding one flux quantum inside the antidot, measuring non-destructively the state of the anyon 
and starting over until the interference  suppression is measured and then we can be sure that the anyon localized 
on the anyon is non-Abelian. In this way we can arrange a topological quantum register of $n$ Fibonacci anyons.
We can use the same type of interferometric measurement to determine the overall topological charge of the register, i.e., 
whether  the register of $n$ Fibonacci anyons belongs to the CFT representation with $|\Lambda\ra =|\I\ra$  or with $|\Lambda\ra =|\varepsilon\ra$
as discussed in Sect.~\ref{sec:single-qubit}.
%
\subsection{Measurement of the Fibonacci qubit}
This is the same non-destructive measurement like in the previous Subsection, however this time applied to pairs of Fibonacci 
anyons localized on neighboring antidots.
Using our encoding scheme in Eq.~(\ref{4F-comp-basis}), if the qubit is in the state $|0\ra$, then the two $\varepsilon$ fields are in the fusion channel of the identity $\I$ 
so there will be no interference suppression.
However, if the  qubit is in the state $|1\ra$, i.e., the two $\varepsilon$ fields are in the fusion channel of  $\varepsilon$ 
then there will be  suppression of the interference term by a factor of $\delta^2$. 
%
\subsection{$n$-qubit encoding scheme}
\label{sec:n-qubit}
\begin{figure}[htb]
	\centering
	\includegraphics[viewport=30 300 550 530,clip,width=\textwidth]{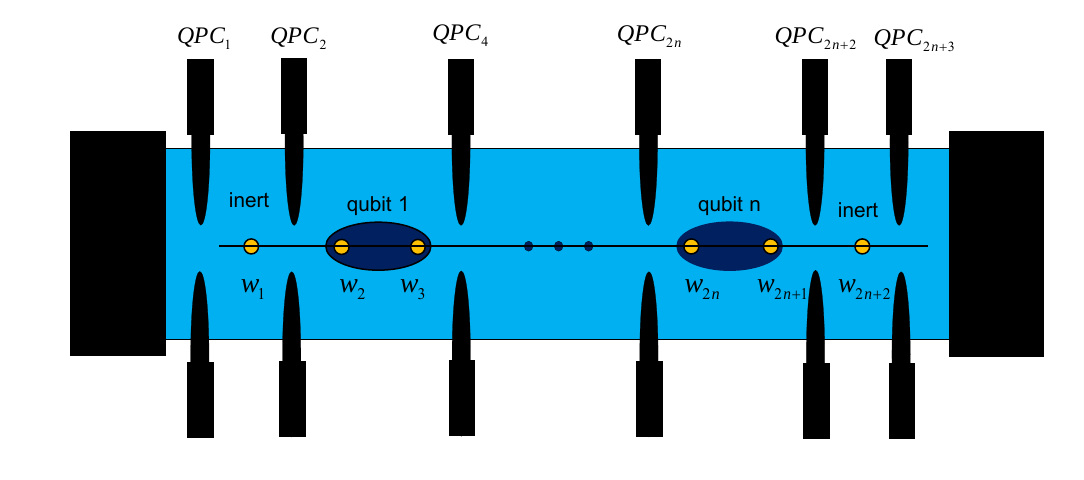} 
\caption{$n$-qubit encoding scheme with $2n+2$ Fibonacci anyons. The yellow disks denote the Fibonacci anyons with 
coordinate $w_i$ localized on antidots  (not shown). There are QPCs between each two neighboring anyons, which can be used for manipulation, 
however those between the anyons forming a qubit (dark blue ellipses) are not shown. The first and the last anyons are inert and the information 
is encoded into   the $n$ pairs of anyons between them.
 \label{fig:n-qubits}}
\end{figure}

In Topological Quantum Computation we intend to encode information in the fusion channels of pairs of Non-Abelian 
anyons \cite{kitaev-TQC,preskill-TQC,sarma-RMP}. Because the Fibonacci anyons $\varepsilon$ have two fusion channels the 
encoding scheme could be the same as that for the Ising anyons \cite{sarma-freedman-nayak,TQC-NPB}. In this Subsection 
we will show that it is possible to construct a $n$-qubit topological register with $2n+2$ Fibonacci anyons (\ref{field-eps}), i.e., 
each qubit is realized by a pair of anyons. Each anyon is localized on an antidot by the flux threading procedure described in
 Sect.~\ref{sec:ini}.
The state of each qubit is determined by the fusion channel of the pair of Fibonacci anyons, i.e., if the pair fuses to the identity 
$\I$ the state of the qubit is
$|0\ra$ while if the pair fuses to the Fibonacci anyon $\varepsilon$ the state of the qubit is $|1\ra$.
Furthermore, because we would like to represent the multi-anyon coordinate wave function as a CFT correlation function as described in 
Sect.~\ref{sec:single-qubit} we need to include one more pair of anyons in order to guarantee that the CFT correlators are non-zero, 
just like in the Ising TQC.
This  encoding scheme is shown in Fig.~\ref{fig:n-qubits}.
The register is arranged so that the first and the last anyons are ``inert'' because they act over the CFT vacuum (left and right)
and create definite states under the state-field correspondence \cite{CFT-book}.
Then the first qubit is formed by the anyons with coordinates $w_2$ and $w_3$.
The $n$-th qubit is formed by the anyons with coordinates $w_{2n}$ and $w_{2n+1}$ and the last anyon carries no information. 
There are QPCs between each two neighboring qubits so that the state of any 
qubit can be measured by
the diagonal conductance interferometry as discussed in Sect.~\ref{sec:FPI}.

The computational basis for $n$-qubits can be expressed in terms of the CFT correlation functions (or more precisely, the conformal blocks)
 with $2n+2$ fields $\varepsilon$, with coordinates $w_i$ ($i=1,\ldots 2n+2$) and
$3r$ parafermion fields $\psi_1$ (which we will skip from the expression)
\beqa\label{n-qubit}
&&|\alpha_1 \alpha_2 \ldots \alpha_n \ra = \nn 
&& \la \varepsilon| \varepsilon(w_2) \mathrm{\Pi}_{\alpha_1}  \varepsilon(w_3) \Pi_1 \varepsilon(w_4) 
\mathrm{\Pi}_{\alpha_2}\varepsilon(w_5)  \Pi_1  \cdots \Pi_1 
\varepsilon(w_{2n}) \Pi_{\alpha_n}  \varepsilon(w_{2n+1})|\varepsilon \ra,
\eeqa
where $\alpha_i=0$ corresponds to the fusion channel $\varepsilon \times \varepsilon = \I$, while $\alpha_i= 1$ corresponds to the
fusion channel $\varepsilon \times \varepsilon = \varepsilon$.
The rule is that every even in order projector \cite{HG-Monodromy} is $\Pi_1$, which corresponds to the odd numbers of $\varepsilon$ in the Bratteli 
diagrams, such as those in Fig.~\ref{fig:4-fibonacci} and  Fig.~\ref{fig:6-fibonacci}. 
Graphically the Bratteli diagrams corresponding to the computational basis
include two consecutive horizontal lines (double ropes) on level $\varepsilon$ and two consecutive ropes (pails) one from level 
$\varepsilon$ to level $\I$ and the second from level $\I$ to level  $\varepsilon$. The notation of the computational basis states is the 
following: from left to right we write $0$ for a pail and $1$ for double ropes. The rest of the CFT blocks which contain odd number of 
horizontal ropes at level $\varepsilon$ are Non-Computational states. The logic behind this rule is that we look for a scalable qubit 
encoding scheme for any $n$. Since one qubit in the state $|1\ra$ is formed by 2 horizontal ropes on level $\varepsilon$, as can be seen from 
Fig.~\ref{fig:4-fibonacci}, there is no way to obtain computational states with odd number of horizontal ropes for multiple qubits.

We shall assume that we can implement the quantum gates by braiding of Fibonacci anyons. The practical execution of the braiding is 
beyond the scope of this paper. What is important for us is that the elementary generators $B_i$, ($i=1,\ldots, n-1$) representing the 
braiding of the 
$i$-th anyon with $i+1$-th can be derived from the coordinate wave functions and must satisfy the Artin's relations for the generators of $\B_n$
\beqa \label{artin}
	B_i B_j &=&  B_j B_i, \qquad \qquad \mathrm{for}  \quad |i-j|\geq 2 \nn
  B_i B_{i+1} B_i &=&  B_{i+1} B_i B_{i+1}, \quad \mathrm{where}
	\quad B_i=R_{i,i+1} \in \B_n.
\eeqa
We shall start in the next section with the one-qubit gates.
\section{Single-qubit gates}
\label{sec:braid}
As we have seen in Eq.~(\ref{4F-comp-basis}) the dimension of the one-qubit space realized in terms of CFT correlators with 4 
$\varepsilon$ fields is 2, as it should be for the one-qubit system. In Eq.~(\ref{decomposed}) we have constructed the 4-anyon wave 
function in terms of the parafermion CFT correlation function with 4 $\varepsilon$ fields and $N=3r$ electron fields with $r$ integer.
The $\uu$ factors of the wave function are of Laughlin type and are written explicitly.

The remaining parafermion CFT correlation function in Eq.~(\ref{decomposed}) was computed in Ref.~\cite{ardonne-schoutens}. 
Following the explicit construction in Ref.~\cite{NPB2001}  of the quasiparticle wavefunctions for the $\Z_k$ parafermion FQH states 
as symmetrization of Laughlin quasiparticle wavefunctions in an Abelian parent CFT the authors of Ref.~\cite{ardonne-schoutens} 
were able to compute the parafermion wave function 
with 4 $\varepsilon$ fields in terms of the hypergeometric function using the original results of Knizhnik and Zamolodchikov \cite{knizh-zam}.
This 4 $\varepsilon$ CFT correlation function is not appropriate for our physical implementation of the Fibonacci TQC because as we can see
in Eq.~(\ref{decomposed}) there must be also $N=3r$ ($r$ is a big positive integer) parafermion fields $\psi_1$. Using the approach of 
Refs.~\cite{NPB2001} and \cite{ardonne-schoutens} the result for general $r$
has been derived in Ref.~\cite{HG-Monodromy} and is explicitly shown in Eqs.~(\ref{Psi_01})--(\ref{Psi_1234}) above.

Now that we have the explicit coordinate wave functions for 4 Fibonacci anyons (\ref{field-eps}) and $N$ electron fields (\ref{psi_h})
we can derive the braid generators $B_a^{(4)}$ by analytic continuation $ (w_a-w_{a+1})\to e^{i\pi }(w_a-w_{a+1})$ of the wave functions. 
The result is obtained in  Ref.~\cite{HG-Monodromy} and we summarize it as follows:
\beq\label{B4}
B_1^{(4)} =\left[\matrix{q^{-1} & 0 \cr 0 &-q}  \right], \quad B_2^{(4)} =\left[\matrix{q^{-3}\t & \sqrt{\t}\cr \sqrt{\t} &-q^3 \t}  \right],
\quad B_3^{(4)} =B_1^{(4)} 
\eeq
where $q=e^{i\pi/5}$ and $\t=1/\delta$ is the inverse of the golden ratio $\delta$ defined after Eq.~(\ref{fusion-fibonacci}).
Notice that our braid generators for 4$\varepsilon$ functions differ from those of Ref.~\cite{HG-Monodromy} by a factor of $q^{3}$ 
which comes from the Laughlin-type factors $(w_a-w_b)^{3/5}$ in Eq.~(\ref{decomposed}) after making the analytic continuation 
$(w_a-w_b)\to e^{i\pi }(w_a-w_b)$ when braiding $w_a$ with $w_b$. It is not difficult to check  \cite{HG-Monodromy} that the braid generators 
(\ref{B4}) of the braid group $\B^{(4)}$ satisfy the Artin's relations (\ref{artin}).

It is important to emphasize one of the findings in Ref.~\cite{ardonne-schoutens} which has been confirmed in Ref.~\cite{HG-Monodromy}: 
it follows from the explicit form of the 4$\varepsilon$ 
function that the 3-point correlation functions of the $\varepsilon$ is zero, i.e., the operator product expansion coefficient is 
$C_{\varepsilon\varepsilon}{}^\varepsilon=0$. However, the operator product expansion of the $\varepsilon$ field is still non-Abelian
\beqa\label{OPE}
\varepsilon(w_1)\varepsilon(w_2) & \mathop{\simeq}_{w_1\to w_2} & \frac{1}{w_{12}^{4/5}} + 
\frac{0}{w_{12}^{2/5}} \varepsilon(w_2)
+\sqrt{\frac{12C}{7}} w_{12}^{3/5} \varepsilon'(w_2), \nn
 2C&=&\sqrt{\frac{\Gamma(\frac{1}{5})\Gamma^3(\frac{3}{5})}{\Gamma(\frac{4}{5})\Gamma^3(\frac{2}{5})}}
\eeqa
where  $ \varepsilon'(w_2)$ is another Virasoro primary field with CFT dimension $\Delta'=7/5$.
This field $\varepsilon'$ is in the same topological sector as $\varepsilon$, which is described by the parfermion character 
$\ch(\L_1+\L_2)$, has the same electric properties however  $\varepsilon'$  is already a true Fibonacci anyon.
This peculiarity does not change our encoding scheme, because we encode information in the fusion channel of pairs of non-Abelian 
anyons--whether the anyons fuse to $\varepsilon$ or to $\varepsilon'$ is not important. What is important is that there are two fusion 
channels - one with quantum dimension 1 and one with quantum dimension $\delta$ and this allows us to construct a qubit. 
From the point of view of the qubit initialization  $\varepsilon'$ is not a problem 
because whenever we create  $\varepsilon$ on the antidot the field  $\varepsilon'$ is created as well.

The generators $B_1^{(4)}$ and $B_3^{(4)}$ of the braid group $\B^{(4)}$ are diagonal and therefore they can be obtained directly 
from the OPE (\ref{OPE}). The generator $B_2^{(4)}$ is not diagonal in the basis $(\Phi^{(0)}, \Phi^{(1)})$ defined in 
Eqs.~(\ref{4e-eta0}) and (\ref{4e-eta1}), however , as has been proven in Ref.~\cite{HG-Monodromy} this matrix is diagonal in the 
\textit{dual basis}  $(\Theta^{(0)}, \Theta^{(1)})$ , which determines the behavior of the hypergeometric functions when 
$\eta \sim 1$ (or, $1-\eta \sim 0$).
Remarkably enough, the two bases are simply related
\beq\label{Theta}
\left[\matrix{\Theta^{(0)} \cr \Theta^{(1)}}\right] = F \left[\matrix{\Phi^{(0)} \cr \Phi^{(1)}}\right], \quad \mathrm{where} \quad
F=\left[\matrix{\tau & \sqrt{\tau} \cr  \sqrt{\tau}  & -\tau}\right] 
\eeq
is the $F$ matrix of Preskill (without any phase on the off-diagonal) \cite{preskill-TQC}. The result in Eq.~(\ref{Theta}) obtained 
in Ref.~\cite{HG-Monodromy} uses the well-known transformation properties of the hypergeometric functions  \cite{bateman-erdelyi}.
Summarizing the results of Ref.~\cite{HG-Monodromy} the braid generator $B_2^{(4)}$ is diagonal in the dual basis (\ref{Theta})
and has the same eigenvalues as $B_1^{(4)}$, i.e., $B_2^{(4)} = F B_1^{(4)} F$. Notice that this result describes the braiding of
4 Fibonacci anyons in the CFT correlation functions including $N\gg 1$ electron fields, generalizing in a non-trivial way 
 the case $N=0$ derived in Ref.~\cite{ardonne-schoutens}.

It is known that the representations of the type (\ref{B4}) of the braid group $\B^{(4)}$ for braiding 4 Fibonacci anyons are dense 
within the unitary group $U(2)$ \cite{preskill-TQC,sarma-RMP}. This means that any single-qubit gate can be implemented by 
products of  $B_i^{(4)}$ and their inverses with an arbitrary precision.
Using the Solovey--Kitaev algorithm \cite{nielsen-chuang} the Hadamard gate can be constructed approximately with 30 weaves with an error 
less than 0.00657 \cite{Hormozi-TQC,H-approx}.

\beqa\label{H}
&&\tilde{H}=B_1^4 B_2^{-2} B_1^2 B_2^{-2} B_1^2 B_2^2 B_1^{-2} B_2^4 B_1^2 B_2^{-2}B_1^{-2}B_2^2 B_1^2= \nn
&&\frac{i}{\sqrt{2}}\left[\matrix{ 1.00402+0.00555 i & 0.99594-0.004772 i \cr 
0.995937+0.00477 i  & -1.00402+0.00555 i}\right] \simeq \, i H,
\eeqa
where we have skipped the superscript $(4)$ of $B_i^{(4)}$ for simplicity, the negative powers of the matrices are the positive powers of the 
inverse matrices, and  for the error  we have used the matrix norm function (i.e., the square root of the maximum eigenvalue of $A A^\dagger$) implemented in Maple
\[
\mathrm{MatrixNorm}(A, 2) = \sqrt{\max(\mathrm{Eigenvalues}(A . A^\dagger)_j, j = 1 , 2)}
\]
for $A=\tilde{H}-i H$.

The Hadamard gate (\ref{H}) together with the exact  single-qubit $Z$ gate
\beq\label{Z}
-Z=\left(B_{1}^{(4)}\right)^5, \quad H Z H=X
\eeq
can be used to approximate to construct an approximation of the single-qubit Pauli-X gate \cite{nielsen-chuang}. Below we give
a better approximation to the $iX$ gate with 44 weaves and accuracy $8.6 \times 10^{-4}$ taken from \cite{bonesteel-2005}
\beqa\label{iX}
\tilde{iX} &=& B_1^{-2}  B_2^{-2}  B_1^2  B_2^{-2}  B_1^2  B_2^{-4}  B_1^2  B_2^4  B_1^{-2}  B_2^4  B_1^{-2}  B_2^2  B_1^2  B_2^{-2}  B_1^4  B_2^{-4} B_1^{-2}  = \nn
&&\left[\matrix{
0.0006 - 0.0002 i  & -0.0006 + 1.     i \cr 
0.0006 + 1.     i  &  0.0006 + 0.0002 i
}\right] \simeq \, iX.
\eeqa
Next, the $T$ gate (or $\pi/8$ gate) \cite{nielsen-chuang} can be approximated upto phase with 14 weaves with accuracy 
$3.3\times 10^{-2}$ as follows
 \beqa\label{T'}
\tilde{T} &=& B_1^{-2}  B_2^1  B_1^{-2}  B_2^1  B_1^{-1}  B_2^2  B_1^{-1}  B_2^1  B_1^{-2}  B_2^1  = \nn
&&\left[\matrix{ 1.     + 0.     i  & 0.0317 + 0.0025 i \cr 
-0.0242 - 0.0207 i  & 0.7055 + 0.7087 i}\right] \simeq \, T.
\eeqa
Some other examples of single-qubit gates are the Pauli $Y$ gate with 45 weaves and accuracy  $8.6 \times 10^{-4}$
 taken from \cite{bonesteel-2005}
 \beqa\label{Y}
\tilde{Y}&=&B_1^2  B_2^2  B_1^{-2}  B_2^2  B_1^{-2}  B_2^4  B_1^{-2}  B_2^{-4}  B_1^2  B_2^{-4}  B_1^2  B_2^{-2}  B_1^{-2}  B_2^2  B_1^{-4}  B_2^4  B_1^{-3}= \nn
&&\left[\matrix{
-0.0006 - 0.0002 i  &  -0.0006 - 1.     i \cr 
-0.0006 + 1.     i  &   0.0006 - 0.0002 i
}\right] \simeq \, Y,
\eeqa
the $-H$ gate (improved over \cite{H-approx}) with 27 weaves and accuracy $4.8 \times 10^{-3}$ taken from \cite{Chinese-H}
\beqa\label{-H}
-\tilde{H} &=& B_1^3  B_2^1  B_1^{-2}  B_2^{-1}  B_1^{-4}  B_2^2  B_1^{-1}  B_2^1  B_1^{-3}  B_2^1  B_1^{-1}  B_2^1  B_1^{-1}  B_2^1  B_1^{-2}  B_2^1  B_1^{-1} = \nn
&&\left[\matrix{
-0.7053 - 0.003  i  & -0.7089 - 0.0029 i \cr 
-0.7089 + 0.0029 i  &  0.7053 - 0.003  i
}\right] \simeq \, -H,
\eeqa
and the $S$ gate multiplied by the phase $q$ with 31 weaves and accuracy $1.2\times 10^{-2}$
\beqa\label{qS}
\tilde{qS} &=& B_2^{-3} B_1^{-4} B_2^2 B_1^3 B_2^{-4} B_1^4 B_2^5 B_1^{-3} B_2^{-3} = \nn
&&\left[\matrix{
0.7085 + 0.7056 i  &  0.0059 - 0.0098 i \cr 
0.0059 + 0.0098 i  & -0.7085 + 0.7056 i
}\right] \simeq \, qS.
\eeqa
Another important exact gate is the $F$ gate in Eq.~(\ref{Theta}) (upto a minus sign) which can be implemented by braiding as follows
\beq
B_1^{(4)}B_2^{(4)}B_1^{(4)} =  -F.
\eeq
%
\section{Two-qubit gates}
\label{sec:Two-qubit}
The two-qubit topological register can be realized with 6 Fibonacci anyons. However, the dimension of 
the Hilbert space with 6 anyons with total quantum dimension 1 is $5$ as can be seen from 
Fig.~(\ref{fig:brattelli-fibonacci}) while the dimension of the two-qubit states is $4$. 
Following the general qubit encoding scheme described in Sect.~\ref{sec:n-qubit} we denote the fusion channel of each pair of Fibonacci anyons as 
a subscript to the parentheses defining the pair and 
 give below the notation for all possible fusion-path states with 6 Fibonacci anyons
\beqa \label{6F-comp-basis}
|00\ra &=& \la \varepsilon| (\varepsilon \varepsilon)_{\I} \   (\varepsilon \varepsilon)_{\I} |\varepsilon \ra\nn
|01\ra &=& \la \varepsilon| (\varepsilon \varepsilon)_{\I} \ (\varepsilon \varepsilon)_{\varepsilon} |\varepsilon \ra\nn
|10\ra &=& \la \varepsilon| (\varepsilon \varepsilon)_{\varepsilon} \ \ (\varepsilon \varepsilon)_{\I} |\varepsilon \ra\nn
|11\ra &=&   \la \varepsilon| (\varepsilon \varepsilon)_{\varepsilon} \ (\varepsilon \varepsilon)_{\varepsilon} |\varepsilon \ra \nn
|NC\ra &=& \la \varepsilon| \varepsilon \ (\varepsilon\varepsilon)_{\I}  \ \varepsilon |\varepsilon  \ra .
\eeqa
Now it is obvious that in addition to the computational basis for two qubits there is another possible fusion-path state which we denote as
$|NC\ra$ as an additional state outside our computational space which we will call a Non-Computational state. What we need for TQC
are the first 4 states
denoted $\left\{  |00\ra ,  |01\ra,   |10\ra,  |11\ra \right\}$. However, as we shall see later, it is possible for some combination of elementary braids 
to arrive at the state  $|NC\ra$ (e.g., when braiding the anyons with coordinates $w_3$ and $w_4$)
which will be leakage of information outside the computational space. This needs special attention when braiding the 6 Fibonacci anyons.

\begin{figure}[htb]
	\centering
	\includegraphics[viewport=100 130 480 710,clip,width=10cm]{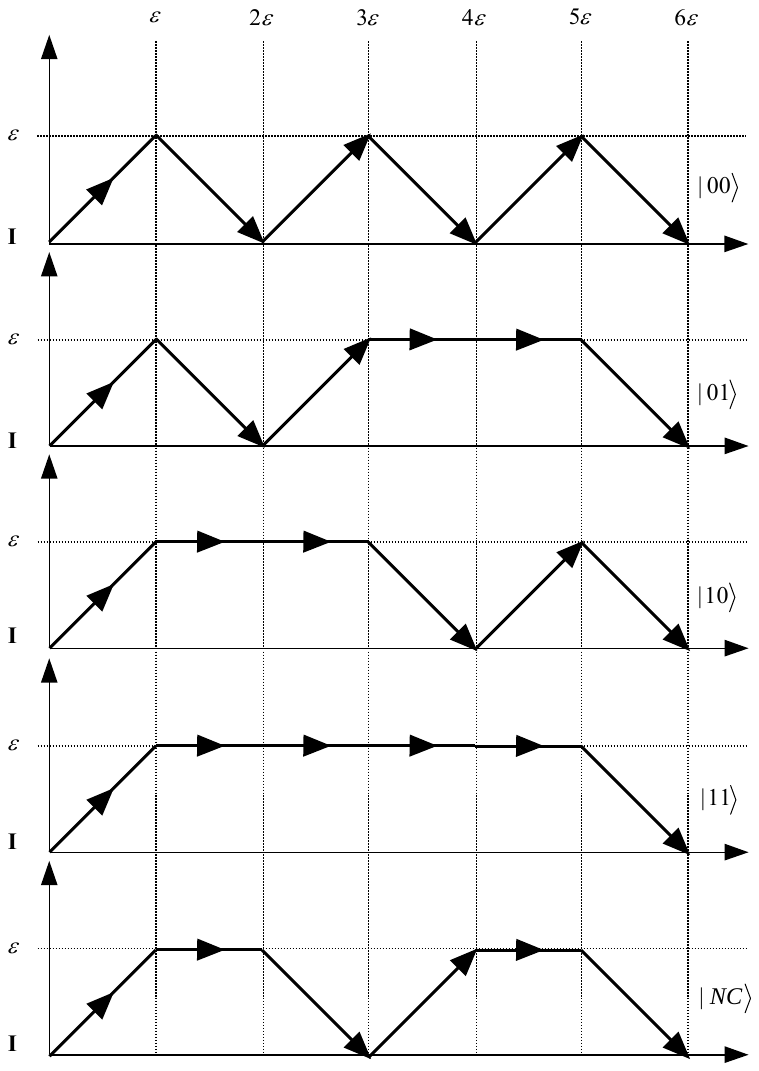} 
\caption{Bratteli diagrams for the 4 states in the computational basis  $\left\{  |00\ra ,  |01\ra,   |10\ra,  |11\ra \right\}$ 
and  the non-computational state $|NC\ra$,  of $6$ Fibonacci anyons with total 
topological charge $\I$ (denoted by \textbf{I} in the figure). 
 \label{fig:6-fibonacci}}
\end{figure}

It is possible and very convenient to represent the fusion-path states by the corresponding Bratteli diagrams, as shown in 
Fig.~\ref{fig:6-fibonacci}. It is 
obvious that when the arrow at the step with 2$\varepsilon$ fields reaches the horizontal axis then the corresponding state of the first qubit is $|0\ra$, 
while if it reaches the dotted line of $\varepsilon$ then the state of the first qubit is $|1\ra$. Similarly, if the composition of the second pair of $\varepsilon$ fields 
reaches at the step with 4$\varepsilon$ fields the horizontal axis then the corresponding state of the second qubit is $|0\ra$, 
while if it reaches the dotted line of $\varepsilon$ then the state of the second qubit is $|1\ra$.
The representation of the generators of the braid group $\B^{(6)}$ 
 have been obtained in Ref.~\cite{HG-Monodromy} 
(in an appropriate basis which differs from the computational one (\ref{6F-comp-basis}) by simple reordering) 
by fusing some of the Fibonacci anyons before braiding, which leads to a direct-sum decomposition into
 into braid generators with $n=4$ (with dimension 2) and $n=5$ (with dimension 3). 
The 5 braid generators of   $\B^{(6)}$ in the basis 
(\ref{6F-comp-basis}) can be written as follows \cite{HG-Monodromy}:
\beqa
B_1^{(6)} &=& \left[ \matrix{q^{-1} & 0 & 0 & 0 &0 \cr 0 & q^{-1} & 0& 0& 0 \cr 0& 0& -q& 0& 0&\cr 0& 0& 0&  -q & 0 \cr 0& 0& 0& 0& -q}\right], \ 
B_2^{(6)} = \left[ \matrix{B_{11} & 0 & B_{12} & 0 &0 \cr  0 &B_{11} & 0 &  B_{12} & 0 \cr  
B_{21} & 0 & B_{22} & 0 &0 \cr 0 & B_{21} & 0 &  B_{22} & 0\cr 0& 0 & 0 & 0 & -q  }\right] \nn
B_3^{(6)}&=&\left[ \matrix{q^{-1} & 0 & 0 & 0 &0 \cr 0& -q & 0&  0& 0 \cr 0 & 0 &   -q & 0 & 0 \cr  
0 & 0 & 0 & B_{11}  & B_{12}  \cr 0 & 0& 0 & B_{21} &  B_{22}}\right] , \
\ \ B_4^{(6)}=\left[ \matrix{ B_{11}  & B_{12} & 0& 0 & 0 \cr  B_{21} &  B_{22} & 0  & 0 & 0  \cr 0 & 0 & B_{11}  &  B_{12} & 0     \cr   
  0 & 0 & B_{21} & B_{22}  & 0 \cr 0&  0 &  0 & 0&  -q }\right] ,  \ \nn
  B_5^{(6)}&=&\left[ \matrix{q^{-1} & 0 & 0 & 0 &0 \cr 0&  -q& 0& 0 & 0&\cr 0 & 0 &  q^{-1} & 0& 0 \cr  0& 0& 0&  -q & 0 \cr 0& 0& 0& 0& -q}\right], \ 
\eeqa
where  $B_{ij}$ with $i,j=1,2$ are the matrix elements of $B_2^{(4)}$ defined in Eq.~(\ref{B4}). 

Obviously, the braid generator  $B_3^{(6)}$ mixes the state $|11\ra$ from the computational basis with the non-computational state 
$|NC\ra$ which may lead to information leakage and loss of unitarity. One  possible solution to the problem of potential information leakage is one or 
two of the $\varepsilon$ fields, e.g., the inert anyons to be excluded from 
the braiding in order to reduce the space dimension  \cite{Hormozi-TQC}.

Notice that the matrices $R:=B_1^{(4)}$ and $B:=B_2^{(4)}$ which can be used to approximate with arbitrary precision
 any operation from $SU(2)$ for a single qubit can be embedded into the two-qubit system by
\beqa
B_1^{(6)}= (R \otimes \I_2)\oplus (-q), \quad
B_2^{(6)} =  (B \otimes \I_2)\oplus (-q) , \nn
B_4^{(6)} =  (\I_2\otimes B)\oplus (-q) , \quad
B_5^{(6)} =  (\I_2 \otimes R)\oplus (-q) .
\eeqa
This means that any single-qubit operation can be implemented in exactly the same way in the two-qubit system.
This also means that 4 of the 5 generators of the group $\B^{(6)}$  are not entangling, because they act as $\I_2$ on one of the qubits, 
and the only entangling generator might be $B_3^{(6)}$.

Two important exact two-qubit gates are the embeddings of the single-qubit $F$ gate into a two-qubit system
\beqa
F_1&=& B_1^{(6)}B_2^{(6)}B_1^{(6)} = -(F \otimes \I_2)\oplus q^{-3} \nn
F_2&=& B_5^{(6)}B_4^{(6)}B_5^{(6)} = -(\I_2\otimes F)\oplus q^{-3} 
\eeqa


\section{Three-qubit gates}
\label{sec:Three-qubit}
\begin{figure}[htb]
	\centering
	\includegraphics[viewport=70 30 530 800,clip,width=10cm]{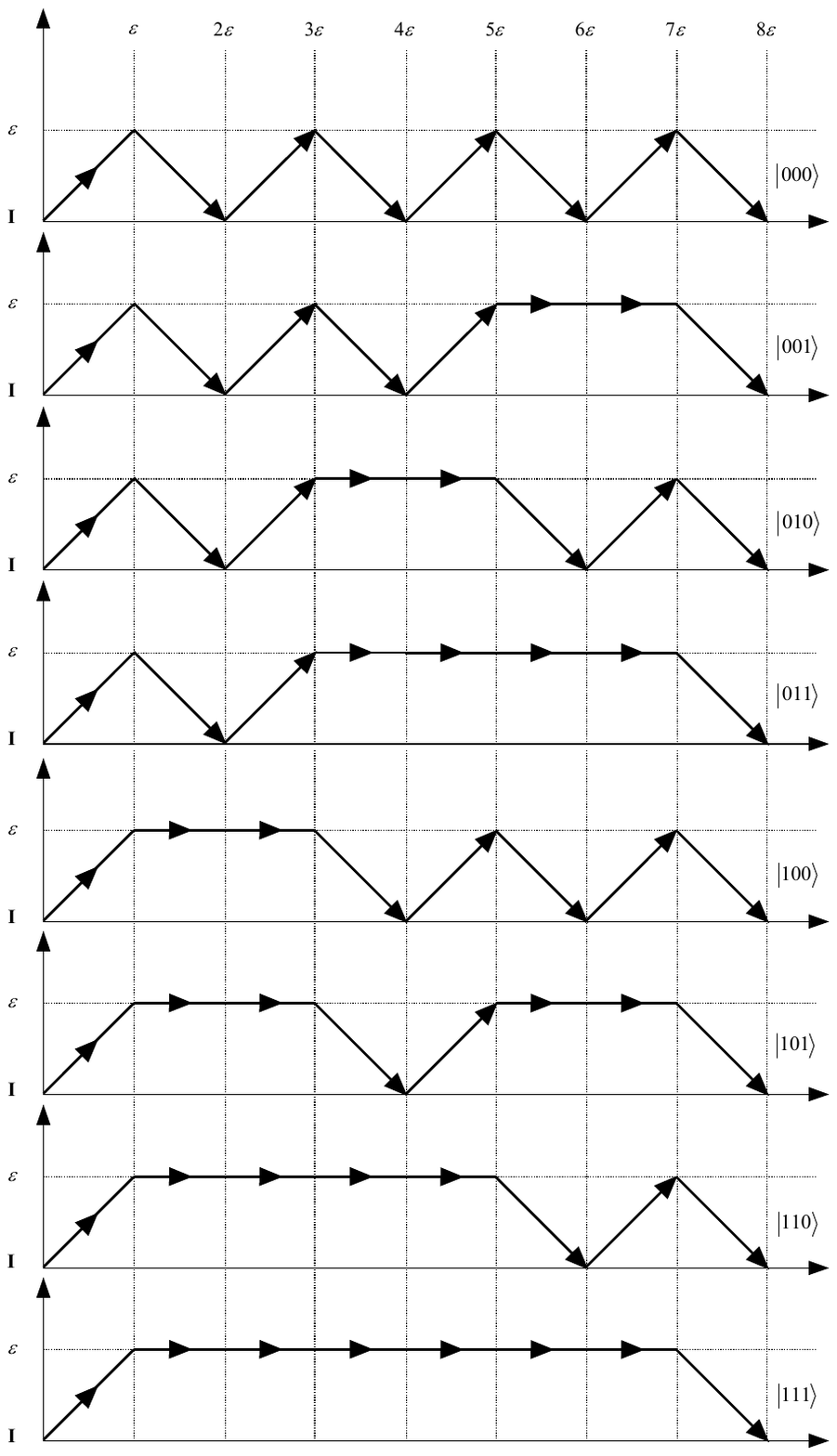} 
\caption{Bratteli diagrams for the 8 states in the computational basis  $\left\{  |000\ra ,  |001\ra,   |010\ra,  |011\ra,  |100\ra ,  |101\ra,   |110\ra,  |111\ra \right\}$ 
  of $8$ Fibonacci anyons with total topological charge $\I$ (denoted by \textbf{I} in the figure). 
 \label{fig:8-fibonacci}}
\end{figure}

\begin{figure}[htb]
	\centering
	\includegraphics[viewport=40 130 540 700,clip,width=10cm]{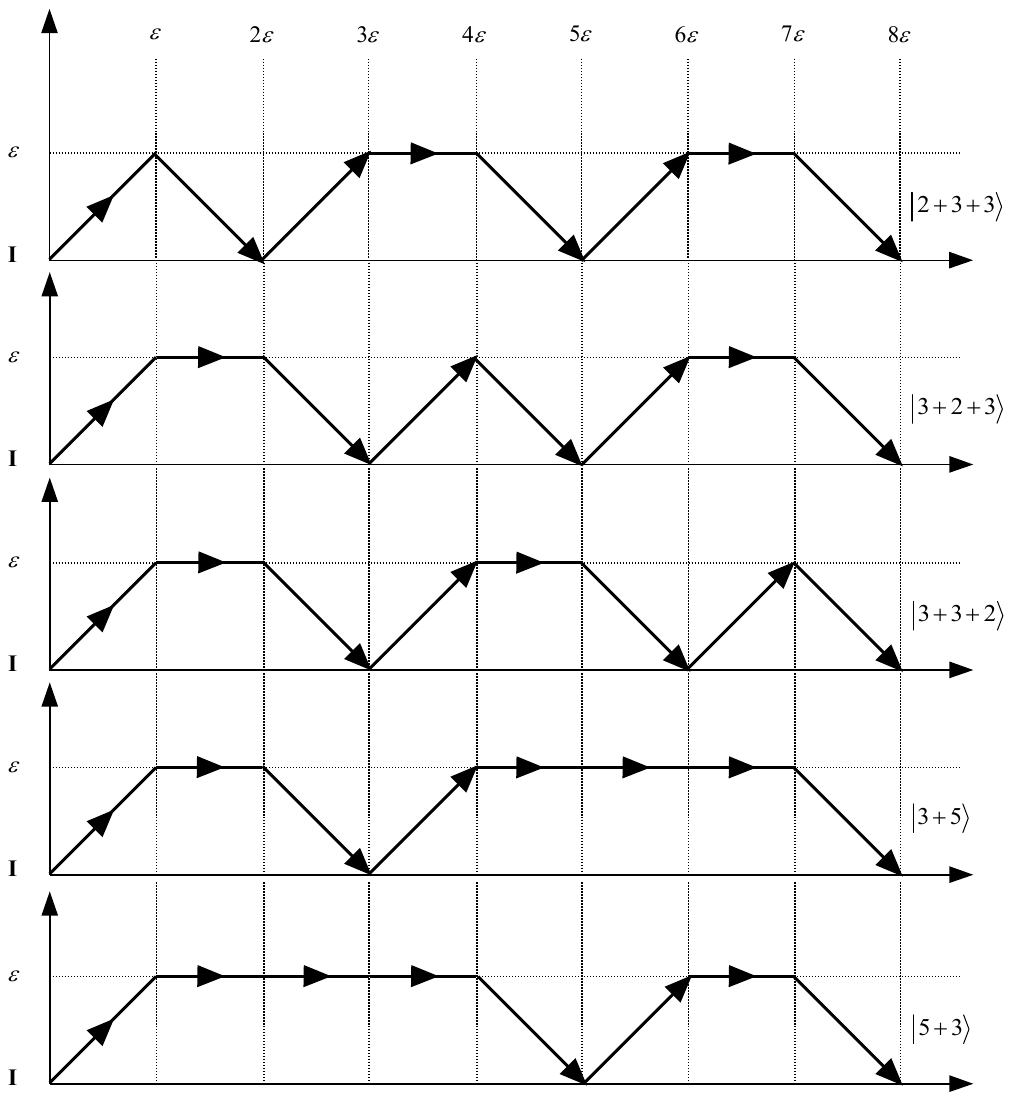} 
\caption{Bratteli diagrams for the 5 non-computational states  of $8$ Fibonacci anyons with total topological charge $\I$  
(denoted by \textbf{I} in the figure)
labeled by the different possible ways to represent the number $8$ as sums of even and odd numbers bigger than 1. 
 \label{fig:8-fibonacci-NC}}
\end{figure}

The topological  register containing three qubits should have dimension $8$ and, at first glance, could eventually be realized with 
7 Fibonacci anyons because the dimension of the fusion-path states is exactly 8. However, following the construction of one and two qubits
by encoding the information in the fusion channel of pairs of Fibonacci anyons it is necessary to use 8 anyons fusing altogether to the vacuum.
Thus the quantum information encoding is as follows: if the $l$-th pair of Fibonacci anyons fuse to $\I$ 
(i.e., the fusion path at position $2l \varepsilon$ touches the horizontal axis in Figure~\ref{fig:8-fibonacci})
then the corresponding quantum bit is $|0\ra$, while if they fuse to the $\varepsilon$ (i.e., the fusion path at position  $2l \varepsilon$ 
touches the $\varepsilon$ 
dashed line in Figure~\ref{fig:8-fibonacci})  then the corresponding quantum bit is $|1\ra$. 
However, the space of the  fusion-path states for wavefunctions containing 8 Fibonacci anyons is bigger than the space spanned by the 
states shown in Figure~(\ref{fig:8-fibonacci}).
Its dimension is $13$ as can be seen from Fig.~\ref{fig:brattelli-fibonacci} and hence there are 5 additional basis vectors which will not participate in the TQC scheme 
and we will again call  them non-computational, see  Fig.~\ref{fig:8-fibonacci-NC} where their fusion-paths are shown.
As can be seen from  Fig.~\ref{fig:8-fibonacci} the three-qubit computational basis can also be labeled by the different ways to represent the number 8 as
 sums of even positive integers bigger that 1.  For example the first computational state $|000\ra$  can be labeled as $|2+2+2+2\ra$ because 
the horizontal  lengths of  the fusion paths connecting two successive points on the horizontal axis  are $2+2+2+2=8$ and the order of the paths is important.

Similarly the non-computational states can be labeled by the different possible ways to represent the number $8$ 
as sums of even and odd numbers bigger than 1, see  Fig.~\ref{fig:8-fibonacci-NC}. For example the first non-computational state is labeled by $2+3+3$ because 
the lengths of  the fusion paths connecting two successive points on the horizontal axis  are $2+3+3=8$ and the order of the paths are important.
It is worth emphasizing that all non-computational states in the Bratteli diagrams  in Fig.~\ref{fig:8-fibonacci-NC}, as well as the two-qubit state $|NC\ra$,  
contain  an odd number of horizontal arrows along the doted horizontal line at level $\varepsilon$ which is impossible to achieve in the qubit encoding 
scheme given in Fig.~\ref{fig:4-fibonacci}. This is why all these states are non-computational.

The generators of the braid  group $\B^{(8)}$ could be expressed  recursively  in terms of the generators of the braid group $\B^{(6)}$ and $\B^{(7)}$.
To see this recall that if we fuse the last two $\varepsilon$ fields we obtain either $\I$ or $\varepsilon$. Therefore the correlation function becomes either 
a topological register with 6 anyons or such a register with 7 anyons.  
 In order to derive them we will use a more convenient ordering of the 8 computational
states shown in Fig.~\ref{fig:8-fibonacci} and 5 non-computational states shown in Fig.~\ref{fig:8-fibonacci-NC}, i.e., 
\beqa\label{Phi}
&&|000\ra, |100\ra, |3+3+2\ra, |010\ra, |110\ra, |3+2+3\ra, |2+3+3\ra,  \nn
&&|5+3\ra,|001\ra, |101\ra, |3+5\ra, |011\ra, |111\ra .
\eeqa
This ordering is such that, if we fuse the last two Fibonacci anyons,  the first 5 states form a representation of the $n=6$ while the second 8 states form a 
representation of $n=7$.
Now it is not difficult to find that all braid generators in the basis (\ref{Phi}), except for the last two, are direct sums of the generators of $\B^{(6)}$ and $\B^{(7)}$, i.e.,
\beq
B_{i}^{(8)}= B_{i}^{(6)}\oplus B_{i}^{(7)}, \quad i=1, \ldots, 5,
\eeq
which is a remarkable manifestation of the Fibonacci numbers, that are the dimensions of the corresponding braid generators. 
The last two generators are explicitly calculated in Ref.~\cite{HG-Monodromy} and can be written in the basis (\ref{Phi}) as follows
\[
B_{6}^{(8)} = \left[ \matrix{ B_{11} \I_{5} & 0 & B_{12} \I_5 \cr 0 & -q \I_3 & 0 \cr B_{21} \I_5 & 0 & B_{22} \I_5}\right], \quad 
B_{7}^{(8)} = \left[ \matrix{q^{-1} \I_{5} & 0 \cr 0 & -q \I_{8} }\right],
\]
where $\I_3$, $\I_5$ and $\I_8$ are the unit matrices with dimensions 3, 5 and 8, respectively,  and $B_{ij}$ with $i,j=1,2$ are the matrix elements of $B_2^{(4)}$
defined in Eq.~(\ref{B4}). This recursive derivation of the generators of the braid group $\B^{(8)}$ is generalized for  $n$ Fibonacci anyons in 
Ref.~\cite{HG-Monodromy} where the generators of the braid groups $\B^{(3)}$, $\B^{(5)}$, $\B^{(7)}$, etc. are also explicitly calculated.

\section*{Acknowledgements}
LSG has been supported as a Research Fellow by the Alexander von Humboldt Foundation. LSG and LH have been supported 
by the Bulgarian Scientific Fund under Contract No. DN 18/3 (2017). This work has been done under the project 
BG05M2OP001-1.002-0006 ''Quantum Communication,  Intelligent Security Systems and Risk Management (QUASAR)'' financed by the
 Operational Program SESG.\\

\bibliography{FQHE,Z_k,my,TQC,QI}

\providecommand{\href}[2]{#2}\begingroup\raggedright\begin{thebibliography}{10}

\bibitem{QTech}
J.~P. Dowling and G.~J. Milburn, ``Quantum technology: the second quantum
  revolution,'' {\em Phil. Trans. R. Soc.} {\bf A 361} (2003) 1655–1674.

\bibitem{QTech-book}
L.~Jaeger, {\em The Second Quantum Revolution}.
\newblock Copernicus Cham, 2018.

\bibitem{nielsen-chuang}
M.~Nielsen and I.~Chuang, {\em Quantum Computation and Quantum Information}.
\newblock Cambridge University Press, 2000.

\bibitem{supremacy-2019}
A.~F., A.~K., and R.~e.~a. Babbush, ``Quantum supremacy using a programmable
  superconducting processor,'' {\em Nature} {\bf 574} (2019) 505–510.

\bibitem{supremacy-2021}
Y.~W. et~al., ``Strong quantum computational advantage using a superconducting
  quantum processor,'' {\em Phys. Rev. Lett.} {\bf 127} (2021) 180501.

\bibitem{kitaev-TQC}
A.~Kitaev, ``Fault-tolerant quantum computation by anyons,'' {\em Ann. of Phys.
  (N.Y.)} {\bf 303} (2003) 2.

\bibitem{preskill-TQC}
J.~Preskill, ``Topological quantum computation,'' {\em Lecture Notes for
  Physics 219} (2004)
  \href{http://xxx.lanl.gov/abs/http://www.theory.caltech.edu/{$\sim$}preskill/ph219}{{\tt
  http://www.theory.caltech.edu/{$\sim$}preskill/ph219}}.

\bibitem{sarma-RMP}
S.~D. Sarma, M.~Freedman, C.~Nayak, S.~H. Simon, and A.~Stern, ``{Non-{A}belian
  Anyons and Topological Quantum Computation},'' {\em Rev. Mod. Phys.} {\bf 80}
  (2008) 1083, \href{http://xxx.lanl.gov/abs/arXiv:0707.1889}{{\tt
  arXiv:0707.1889}}.

\bibitem{simon-TQ}
S.~H. Simon, {\em Topological Quantum}.
\newblock Oxford University Press, Oxford UK, 2023.

\bibitem{wilczek}
F.~Wilczek, {\em Fractional statisitcs and anyon superconductivity}.
\newblock World Scientific, Singapore, 1990.

\bibitem{mr}
G.~Moore and N.~Read, ``Nonabelions in the fractional quantum {H}all effect,''
  {\em Nucl. Phys.} {\bf B360} (1991) 362.

\bibitem{5-2-book}
K.~K.~W. Ma, M.~R. Peterson, V.~W. Scarola, and K.~Yang, {\em Encyclopedia of
  Condensed Matter Physics}, vol.~1, ch.~Fractional quantum Hall effect at the
  filling factor $\nu=5/2$, pp.~324--365.
\newblock Elsevier, 2nd~ed., 2024.

\bibitem{clifford}
A.~Ahlbrecht, L.~S. Georgiev, and R.~F. Werner, ``Implementation of {C}lifford
  gates in the {Ising}-anyon topological quantum computer,'' {\em Phys. Rev.}
  {\bf A 79} (2009) 032311, \href{http://xxx.lanl.gov/abs/arXiv:0812.2338}{{\tt
  arXiv:0812.2338}}.

\bibitem{choi-west-08}
H.~C. Choi, W.~Kang, S.~D. Sarma, L.~N. Pfeiffer, and K.~W. West, ``Activation
  gaps of fractional quantum {H}all effect in the second {L}andau level,'' {\em
  Phys. Rev.} {\bf B 77} (2008) 081301.

\bibitem{rr}
N.~Read and E.~Rezayi, ``Beyond paired quantum {H}all states: parafermions and
  incompressible states in the first excited {L}andau level,'' {\em Phys. Rev.}
  {\bf B59} (1998) 8084.

\bibitem{NPB2001}
A.~Cappelli, L.~S. Georgiev, and I.~T. Todorov, ``Parafermion {H}all states
  from coset projections of {A}belian conformal theories,'' {\em Nucl. Phys.}
  {\bf B 599 [FS]} (2001) 499--530,
  \href{http://xxx.lanl.gov/abs/hep-th/0009229}{{\tt hep-th/0009229}}.

\bibitem{freedman-larsen-wang-TQC}
M.~Freedman, M.~Larsen, and Z.~Wang, ``A modular functor which is universal for
  quantum computation,'' {\em Commun. Math. Phys.} {\bf 227} (2002) 605,
  \href{http://xxx.lanl.gov/abs/quant-ph/0001108}{{\tt quant-ph/0001108}}.

\bibitem{bonesteel-2005}
N.~Bonesteel, L.~Hormozi, G.~Zikos, and S.~Simon, ``Braid topologies for
  quantum computation,'' {\em Phys. Rev. Lett.} {\bf 95} (2005) 140503.

\bibitem{Hormozi-TQC}
L.~Hormozi, G.~Zikos, N.~E. Bonesteel, and S.~H. Simon, ``Topological quantum
  compiling,'' {\em Phys. Rev.} {\bf B 75} (2007) 165310.

\bibitem{Hormozi-TQC-RR}
L.~Hormozi, N.~Bonesteel, and S.~Simon, ``Topological quantum computing with
  {Read--Rezayi} states,'' {\em Phys. Rev. Lett.} {\bf 103} (2009) 160501.

\bibitem{sarma-freedman-nayak}
S.~D. Sarma, M.~Freedman, and C.~Nayak, ``Topologically-protected qubits from a
  possible non-{A}belian fractional quantum {H}all state,'' {\em Phys. Rev.
  Lett.} {\bf 94} (2005) 166802,
  \href{http://xxx.lanl.gov/abs/cond-mat/0412343}{{\tt cond-mat/0412343}}.

\bibitem{TQC-PRL}
L.~S. Georgiev, ``Topologically protected gates for quantum computation with
  non-{A}belian anyons in the {P}faffian quantum {H}all state,'' {\em Phys.
  Rev.} {\bf B 74} (2006) 235112,
  \href{http://xxx.lanl.gov/abs/cond-mat/0607125}{{\tt cond-mat/0607125}}.

\bibitem{TQC-NPB}
L.~S. Georgiev, ``Towards a universal set of topologically protected gates for
  quantum computation with {P}faffian qubits,'' {\em Nucl. Phys.} {\bf B 789}
  (2008) 552--590, \href{http://xxx.lanl.gov/abs/hep-th/0611340}{{\tt
  hep-th/0611340}}.

\bibitem{HG-Monodromy}
L.~Hadjiivanov and L.~S. Georgiev, ``Braiding {F}ibonacci anyons,''
  \href{http://xxx.lanl.gov/abs/arXiv:2404.01778}{{\tt arXiv:2404.01778}}.

\bibitem{ardonne-schoutens}
E.~Ardonne and K.~Schoutens, ``Wavefunctions for topological quantum
  registers,'' {\em Ann. Phys.} {\bf 322} (2007) 201--235.

\bibitem{CFT-book}
P.~{Di Francesco}, P.~Mathieu, and D.~S\'en\'echal, {\em Conformal Field
  Theory}.
\newblock Springer--Verlag, New York, 1997.

\bibitem{JGSP-S-PF}
L.~S. Georgiev, ``{Exact Modular $S$ Matrix for $\Z_k$ Parafermion Quantum Hall
  Islands and Measurement of Non-Abelian Anyons},'' {\em J. of Geom. and Symm.
  in Phys.} {\bf 62} (2021) 1--28.

\bibitem{pan}
W.~Pan, J.-S. Xia, V.~Shvarts, D.~E. Adams, H.~L. St\"ormer, D.~C. Tsui, L.~N.
  Pfeiffer, K.~W. Baldwin, and K.~W. West, ``Exact quantization of the
  even-denominator fractional quantum {H}all state at $\nu = 5/2$ {L}andau
  level filling factor,'' {\em Phys. Rev. Lett.} {\bf 83} (1999) 3530,
  \href{http://xxx.lanl.gov/abs/cond-mat/9907356}{{\tt cond-mat/9907356}}.

\bibitem{xia}
J.~Xia, W.~Pan, C.~Vicente, E.~Adams, N.~Sullivan, H.~Stormer, D.~Tsui,
  L.~Pfeiffer, K.~Baldwin, and K.~West, ``Electron correlation in the second
  {L}andau level: a competition between many nearly degenerated quantum
  phases,'' {\em Phys. Rev. Lett.} {\bf 93} (2004) 176809.

\bibitem{pan-xia-08}
W.~Pan, J.~S. Xia, H.~L. Stormer, D.~C. Tsui, C.~Vicente, E.~D. Adams, N.~S.
  Sullivan, L.~N. Pfeiffer, K.~W. Baldwin, and K.~W. West, ``Experimental
  studies of the fractional quantum {H}all effect in the first excited {L}andau
  level,'' {\em Phys. Rev.} {\bf B 77} (Feb, 2008) 075307.

\bibitem{fro-stu-thi}
J.~Fr\"{o}hlich, U.~M. Studer, and E.~Thiran, ``A classification of quantum
  {H}all fluids,'' {\em J. Stat. Phys.} {\bf 86} (1997) 821,
  \href{http://xxx.lanl.gov/abs/cond-mat/9503113}{{\tt cond-mat/9503113}}.

\bibitem{LT9}
L.~S. Georgiev, ``Hilbert space decomposition for {C}oulomb blockade in
  {F}abry--{P}\'{e}rot interferometers,'' in {\em Lie Theory and Its
  Applications in Physics: IX International Workshop}, V.~Dobrev, ed., Springer
  Proceedings in Mathematics \& Statistics 36, pp.~439--450.
\newblock 2011.
\newblock \href{http://xxx.lanl.gov/abs/arXiv:1112.5946}{{\tt
  arXiv:1112.5946}}.
\newblock Proceedings of the 9-th International Workshop "Lie Theory and Its
  Applications in Physics", 20-26 June 2011, Varna, Bulgaria.

\bibitem{NPB2015-2}
L.~S. Georgiev, ``Thermopower and thermoelectric power factor of ${\Z}_k$
  parafermion quantum dots,'' {\em Nucl. Phys.} {\bf B 899} (2015) 289--311,
  \href{http://xxx.lanl.gov/abs/arXiv:1505.02538}{{\tt arXiv:1505.02538}}.

\bibitem{cz}
A.~Cappelli and G.~R. Zemba, ``Modular invariant partition functions in the
  quantum {H}all effect,'' {\em Nucl. Phys.} {\bf B490} (1997) 595,
  \href{http://xxx.lanl.gov/abs/hep-th/9605127}{{\tt hep-th/9605127}}.

\bibitem{NPB-PF_k}
L.~S. Georgiev, ``A universal conformal field theory approach to the chiral
  persistent currents in the mesoscopic fractional quantum {H}all states,''
  {\em Nucl. Phys.} {\bf B 707} (2005) 347--380,
  \href{http://xxx.lanl.gov/abs/hep-th/0408052}{{\tt hep-th/0408052}}.

\bibitem{NPB2015}
L.~S. Georgiev, ``{Thermoelectric properties of Coulomb-blockaded fractional
  quantum Hall islands},'' {\em Nucl. Phys.} {\bf B 894} (2015) 284--306,
  \href{http://xxx.lanl.gov/abs/arXiv:1406.6177}{{\tt arXiv:1406.6177}}.

\bibitem{schweber}
S.~Schweber, {\em An Introduction to Relativistic Quantum Field Theory}.
\newblock Row, Peterson and Company, 1961.

\bibitem{bateman-erdelyi}
H.~Bateman and A.~Erdelyi, {\em Higher Transcendential Functions}, vol.~1.
\newblock McGraw-Hill, New York, 1953.

\bibitem{laughlin}
R.~B. Laughlin, ``Anomalous quantum {H}all effect: An incompressible quantum
  fluid with fractionally charged excitations,'' {\em Phys. Rev. Lett.} {\bf
  50} (May, 1983) 1395--1398.

\bibitem{bonderson-5-2}
P.~Bonderson, A.~Kitaev, and K.~Shtengel, ``Detecting non-abelian statistics in
  the $\nu=5/2$ fractional quantum {H}all state,'' {\em Phys. Rev. Lett.} {\bf
  96} (2006) 016803, \href{http://xxx.lanl.gov/abs/cond-mat/0508616}{{\tt
  cond-mat/0508616}}.

\bibitem{stern-halperin-5-2}
A.~Stern and B.~I. Halperin, ``Proposed experiments to probe the non-{A}belian
  $\nu=5/2$ quantum {H}all state,'' {\em Phys. Rev. Lett.} {\bf 96} (2006)
  016802.

\bibitem{bonderson-12-5}
P.~Bonderson, K.~Shtengel, and J.~K. Slingerland, ``Probing non-abelian
  statistics with two-particle interferometry,'' {\em Phys. Rev. Lett.} {\bf
  97} (2006) 016401, \href{http://xxx.lanl.gov/abs/cond-mat/0601242}{{\tt
  cond-mat/0601242}}.

\bibitem{knizh-zam}
V.~G. Knizhnik and A.~B. Zamolodchikov, ``{Current Algebra and Wess--Zumino
  Model in Two Dimensions},'' {\em Nucl. Phys.} {\bf B 247} (1984) 83--103.

\bibitem{H-approx}
M.~T. Rouabah, N.~E. Belaloui, and A.~Tounsi, ``Compiling single-qubit braiding
  gate for {F}ibonacci anyons topological quantum computation,'' {\em J. of
  Phys: Conference Series} {\bf 1766} (2021) 012029.

\end{thebibliography}\endgroup

\end{document}